\documentclass{aastex}
\usepackage{emulateapj5}
\usepackage{epsfig}

\newcommand{\eq}[1]{equation (\ref{#1})}
\newcommand{\Eq}[1]{Equation (\ref{#1})}
\newcommand{\fig}[1]{Figure {\scshape \ref{#1}}}
\newcommand{\tab}[1]{Table \ref{#1}}
\newcommand{\nbody}{\mbox{$N$-body}}

\makeatletter

\newenvironment{inlinefigure}{%
\def\@captype{figure}%
\noindent\begin{minipage}{0.999\linewidth}\begin{center}}
{\end{center}\end{minipage}\smallskip}
\makeatother

\begin{document}

\shorttitle{MASS SEGREGATION IN GLOBULAR CLUSTERS}
\shortauthors{FREGEAU ET AL.}
\submitted{Accepted for Publication in ApJ}

\title{Mass Segregation in Globular Clusters}

\author{J.\ M.\ Fregeau\altaffilmark{1}, K.\ J.\ Joshi\altaffilmark{2}, 
 S.\ F.\ Portegies Zwart\altaffilmark{3,4}, F.\ A.\ Rasio\altaffilmark{5}}
\altaffiltext{1}{Department of Physics, MIT, 77 Massachusetts Ave, Cambridge, MA 02139; {\it fregeau@mit.edu}}
\altaffiltext{2}{Present address: 75 Peterborough St  \#313, Boston, MA 02215; {\it kjoshi@alum.mit.edu}}
\altaffiltext{3}{Astronomical Institute `Anton Pannekoek', University of Amsterdam, Netherlands; {\it spz@space.mit.edu}}
\altaffiltext{4}{Hubble Fellow}
\altaffiltext{5}{Department of Physics and Astronomy, Northwestern University, 2145 Sheridan Rd,
 Evanston, IL 60208; {\it rasio@northwestern.edu}}

\begin{abstract}
We present the results of a new study of mass segregation in
two-component star clusters, based on a large number of numerical
$N$-body simulations using our recently developed dynamical Monte
Carlo code.  Specifically, we follow the dynamical evolution of 
clusters containing stars with individual masses $m_1$ as well as a
tracer population of objects with individual masses $m_2$.  We
consider both light tracers ($\mu \equiv m_2/m_1 < 1$) and heavy
tracers ($\mu > 1$), and a variety of King model initial conditions.
In all of our simulations we use a realistically large number of stars
for globular clusters, $N=10^5$, but we ignore the effects of binaries
and stellar evolution. For heavy tracers, which could represent
stellar remnants such as neutron stars or black holes in a globular
cluster, we characterize in a variety of ways the tendency for these
objects to concentrate in or near the cluster core.  In agreement with
simple theoretical arguments, we find that the characteristic time for
this mass segregation process varies as $1/\mu$.  For models with very
light tracers ($\mu \lesssim 10^{-2}$), which could represent
free-floating planets or brown dwarfs, we find the expected depletion
of light objects in the cluster core, but also sometimes a significant
{\it enhancement\/} in the halo.  That is, for some initial
conditions, the number density of light objects in the cluster halo
increases over time, in spite of the higher overall evaporation rate
of lighter objects through the tidal boundary.  Using these results
along with a simplified initial mass function, we estimate the optical
depth to gravitational microlensing by planetary mass objects or brown
dwarfs in typical
globular clusters.  For some initial conditions, the optical depth in
the halo due to very low-mass objects could be much greater than that of
luminous stars.  If we apply our results to M22, using the recent null detection
of \citet{sahu2}, we find an upper limit of $\sim 25\%$ at the 63\%
confidence level for the current
mass fraction of M22 in the form of very low-mass objects.
\end{abstract}


\section{Introduction}

Globular clusters are thought to contain objects with a very wide
range of masses, although even the present-day mass function is rather
poorly constrained by observations (see, e.g., Bedin et al.\ 2001).
The initial mass function (IMF) is even less constrained, as it
depends also on the details of the overall cluster dynamical evolution
(Vesperini \& Heggie 1997).  Indeed, even for highly idealized systems
of unevolving point masses, a wide mass spectrum, combined with the
effects of two-body relaxation, leads to a variety of complex
dynamical phenomena that are still poorly understood theoretically
(Heggie et al.\ 1998). We will refer to these phenomena collectively
here as ``mass segregation,'' but note that they involve a number of
different processes such as as mass loss through the tidal boundary,
energy equipartition and mass stratification, gravothermal
contraction, and even, in some cases, gravothermal instabilities (see,
e.g., Spitzer 1987).

The masses of directly observable stars today in globular clusters
cover a relatively narrow range from $\sim1-2\,M_\odot$ (e.g.,
primordial binaries, blue stragglers, neutron stars; see, e.g., Bailyn
1995) down to $\sim 0.1\,M_\odot$ at the faint end of the main
sequence (e.g., Marconi et al.\ 1998).  Much more massive stars were
certainly present earlier in the dynamical evolution of these
clusters, including more massive main-sequence stars and binaries, as
well as primordial $\sim 10\,M_\odot$ black holes (Portegies Zwart \&
McMillan 2000).  Much lower mass objects such as brown dwarfs or
planets may also have formed in large numbers within the cluster
initially and, as we will show in this paper, significant numbers
could also have been retained to the present.

Some lower mass objects $\sim 10^{-2}-10^{-3}\,M_\odot$ have been
detected in globular clusters as companions to millisecond radio
pulsars (D'Amico et al.\ 2001; Ford et al.\ 2000; Freire et al.\
2001).  Gilliland et al.\ (2001) used {\it HST} to search for transits by
giant planets in short-period orbits (``hot Jupiters'') around
main-sequence stars in the central region of 47~Tuc. They reported a
negative result and concluded that the planet frequency in 47~Tuc must
be at least an order of magnitude below that for the solar
neighborhood. However, in the high-density central region of this
cluster, where the search was conducted, planetary systems are likely
to be disrupted by encounters with other stars and binaries (Davies \&
Sigurdsson 2001), and the frequency of ``free-floating'' planets is
not constrained by these observations.  In contrast, gravitational
microlensing can be used to try to detect such ``free-floating''
planets (Paczy\'nski 1994).  Analysis of microlensing events toward
the galactic bulge suggests the presence of a substantial amount
of lower mass objects in globular clusters \citep{jetzer}.  
More recently, in a pioneering study, Sahu et al.\ (2001)
monitored about 83,000 Galactic bulge stars for microlensing
by objects in the globular cluster M22.  They reported one clear
microlensing event associated with an object of about $0.1\,M_\odot$,
and six events unresolved in time that were very tentatively
associated with planetary-mass objects in the cluster

Observational evidence for mass segregation has been found in a number
of globular clusters. In some cases, direct evidence comes from the
measurement of a more centrally concentrated density profile for some
heavier stellar population, such as blue stragglers or radio pulsars 
(C\^ot\'e, Richer \& Fahlman 1991; Layden
et al.\ 1999; Rasio 2000). More often, the evidence comes from an
apparent decrease in the slope of the (continuous) mass function toward
the cluster core.  This can either be estimated from the observed
luminosity function at different radii (Howell et al.\ 2001; Sosin
1997) or inferred from color gradients (Howell, Guhathakurta, \& Tan
2000).  Note that, in contrast to younger star clusters where the
present-day mass segregation may still reflect initial conditions
(Figer et al.\ 1999; but see Portegies Zwart et al.\ 2002), any mass
segregation observed in the central regions of globular clusters must
be a result of dynamical evolution, since the relaxation time in those
regions is typically much shorter than globular cluster ages (by a
factor $\sim 10-10^3$).

Our theoretical understanding of the dynamical evolution of dense star
clusters containing a wide mass spectrum is far from complete, even in
the highly idealized limit of unevolving point masses. Direct $N$-body
simulations are very computationally intensive and have therefore been
limited to systems with unrealistically low $N\sim10^3-10^4$ and
rather narrow mass spectra (de la Fuente Marcos 1996; Giersz \& Heggie
1997; Takahashi \& Portegies Zwart 2000).  Since realistic IMFs are
thought to increase (perhaps steeply) toward smaller masses, a
lower-mass cut-off (usually $\sim 0.1-0.5\,M_\odot$) is introduced to
avoid ending up with very few heavier objects and a large numerical
noise in the simulation (since these heavier objects often dominate
the overall dynamics).

Instead, in this paper, we use H\'enon's {\it Monte Carlo\/} method to
compute the dynamical evolution of clusters containing a more
realistic number of stars ($N=10^5$). Our Monte Carlo code, as well as
a number of test calculations and comparisons with direct $N$-body
integrations, have been described in detail in Joshi, Rasio, \&
Portegies Zwart (2000) and Joshi, Nave, \& Rasio (2001).  As a first
step, we consider in this paper the simplest {\it two-component
clusters\/}, in which only two types of objects, of mass $m_1$ and
$m_2\ne m_1$, are present. In addition, we assume that most objects in
the cluster are of mass $m_1$, while the objects of mass $m_2$ form a
{\it tracer population\/}, i.e., the total component mass ratio
$M_2/M_1 \ll 1$.  In general, the gravothermal evolution of a
two-component cluster can be either stable or unstable, depending on
the two ratios $m_2/m_1$ and $M_2/M_1$ (Spitzer 1969). In the stable
case, the two components remain thermally coupled and the cluster
evolves dynamically toward energy equipartition between the two
species.  In the unstable case, energy equipartition is impossible and
the heavier species decouples thermally from the rest of the cluster,
evolving separately toward core collapse and interacting with the
lighter component through the mean gravitational field only.  In a
previous paper (Watters, Joshi, \& Rasio 2000) we studied
systematically the development of this instability (sometimes called
the Spitzer ``mass-stratification'' instability) in two-component
systems, and we determined the location of the stability boundary in
the parameter space of $m_2/m_1$ and $M_2/M_1$.  Instead, in this
paper, complementary to our previous study, we concentrate on the
stable systems with $M_2/M_1\ll 1$ and we study systematically the
effects of their dynamical evolution toward energy equipartition. In
particular, we seek to characterize quantitatively the tendency for
heavier objects to develop radial density profiles that are more
centrally concentrated, as well as the tendency for lighter objects to
concentrate away from the central regions, and to be preferentially
evaporated in the Galactic tidal field.

Our paper is organized as follows. In Sec.~2 we present an overview of
our numerical approach, and we describe our initial conditions. In
Sec.~3 we present our results for light tracers (with $m_2 < m_1$) and
we discuss their implications for microlensing by low-mass objects in
globular clusters.  In Sec.~4 we present our results for heavy tracers
(with $m_2 > m_1$) and we compare them to simple analytical
estimates. Our conclusions and summary are presented in Sec.~5.

\section{Overview of Numerical Method}

We study exclusively two-component clusters comprised of point masses, 
ignoring the effects of binaries and stellar evolution.
To evolve the clusters, we use our recently developed 2-D Monte Carlo
code \citep{joshi1}, which is based on H\'enon's Monte Carlo algorithm
for solving the Fokker-Planck equation \citep{henon}.  Our code has shown
close agreement with direct {\nbody} and Fokker-Planck calculations of
one-component clusters \citep{joshi1}, as well as calculations incorporating
mass spectra and tidal mass loss \citep{joshi2}.  Furthermore, we have demonstrated
reasonable agreement with the results of many different codes from
Heggie's ``Collaborative Experiment'' (\citet{heggie}; see especially
Fig.~2 of \citet{rasio2001}).  

For the initial conditions, we use the King model (a lowered Maxwellian),
given by the distribution function
\begin{equation}
\label{one}
f = \left\{\kappa \left(e^{{\cal E}/\sigma^2} - 1\right) \quad\hbox{for ${\cal E} > 0$}
	\atop 0 \hfill\hbox{for ${\cal E} \leq 0$}\right. \, ,
\end{equation}
where ${\cal E} \equiv \Psi - {1 \over 2}v^2$ is the (negative) energy of each star relative
to the potential at the tidal radius $r_t$, $\Psi \equiv -\Phi + \Phi(r_t)$ is the relative
gravitational potential,
and $\kappa$ is a normalization constant.  The quantity $\sigma$ is a parameter, and should not 
be confused with the actual velocity dispersion $\sqrt{\langle v^2 \rangle}$.  
See, e.g., \citet{bandt}, or \citet{spitzer87}.
For light tracers we enforce the tidal boundary induced by the Galaxy using a simple spherical
Roche approximation (Joshi et al.\ 2001), while for heavy tracers (which tend to concentrate
near the cluster center and hardly ever get ejected) we treat the clusters as isolated.

To effect a two-component distribution, we first create a
single-component distribution of stars of mass $m_1$ according to
\eq{one}, and then randomly replace a fraction of these background
stars with tracers of mass $m_2$, so that the initial density profile of the
tracers is the same as that of the background.  Although this creates
clusters that are initially slightly out of virial equilibrium, we
find that they settle into virial equilibrium within a few crossing times
(a few timesteps).
We have verified this independently using direct {\nbody\/}
integrations (see below), which relax to virial equilibrium within a
few crossing times.

For light tracers (with mass ratio $\mu \equiv m_2/m_1 < 1$), 
we consider a variety of King models for the initial
conditions, with dimensionless central potentials $W_0 \equiv
\Psi(0)/\sigma^2 = \left\{ 1, 3, 5, 7, 9 \right\}$, a variety of mass
ratios in the range $0.001 \leq \mu \leq 0.4$, and a tracer population
of either $N_t = 1000$ or $N_t = 15000$ out of $10^5$ total stars.  
Table 1 gives the initial conditions for each model.  In all cases, tracers
make up no more than 0.5\% of the mass of each cluster, and
consequently do not significantly affect the overall evolution of the
cluster.

For heavy tracers, we consider King models with $W_0 = \left\{ 3, 7, 10 \right\}$,
a variety of mass ratios in the range
$1.5 \leq \mu \leq 10$, and a tracer population in the range $100 \leq N_t \leq 1000$.  With 
a mass ratio greater than one, it is not always feasible to use such a small number of tracers 
that the cluster is less than a few percent tracers by mass, since then statistical noise is too great.
Instead, we choose a reasonable value for the number of tracers, and consequently explore both
systems which can reach energy equipartition, and systems which evolve away from thermal
equilibrium. 
A simple test to determine whether thermal equilibrium is possible for a two-component
cluster is the Spitzer stability condition \citep{spitzer69},
\begin{equation}
\label{two}
S \equiv \left({M_2 \over M_1}\right)\left({m_2 \over m_1}\right)^{3/2} \lesssim 0.16 \, ,
\end{equation}
where $M_1$ and $M_2$ are the total masses of species 1 and 2, respectively.  If $S$ is 
less than about $0.16$, the system is ``Spitzer-stable,'' and the two components will 
reach thermal equilibrium.  If $S$ exceeds this value, the system is said to be 
``Spitzer-unstable,'' and equipartition is impossible.  \citet{watters} performed a more refined analysis
using our dynamical Monte Carlo code to evolve a wide range of two-component
clusters, and arrived at the more accurate empirical stability criterion
\begin{equation}
\label{three}
\Lambda \equiv \left({M_2 \over M_1}\right)\left({m_2 \over m_1}\right)^{2.4} \lesssim 0.32 \, .
\end{equation}
Table 2 gives these stability parameters, as well as the initial conditions, for 
each model considered.

Some calculations with $\mu < 1$ were repeated with direct {\nbody}
integrations. For these direct {\nbody} calculations, we used the program
{\tt kira} within the
Starlab software environment (see \citet{portegieszwart} and {\tt http://manybody.org})
and used the special-purpose
GRAPE-4 and GRAPE-6 hardware (a single GRAPE-6 board) \citep{makino97} to accelerate the computation of
gravitational forces between stars.  Time integration of stellar orbits was accomplished using a
fourth-order Hermite scheme \citep{makino92}. {\tt kira} also
incorporates block timesteps \citep{mcmillan1,mcmillan2,makino91}
and a special treatment of close two-body and multiple encounters of
arbitrary complexity.
Given the high cost of the direct {\nbody} integrations we limited our
selection of initial conditions to a few models with $\mu < 1$ and performed the 
calculations with $N=6144$ stars; these calculations take of order 10 days each to 
complete on either the GRAPE-4 or a single GRAPE-6 board.  
(With our Monte-Carlo code, each calculation takes $\sim 12$ hours.)

Following the convention of most previous studies, we define dynamical units so that
$G=M_0=-4E_0=1$, where $M_0$ and $E_0$ are the initial total mass and total energy
of the cluster \citep{henon}.  The units of length, $L$, and time, $T$, are then
\begin{equation}
\label{time1}
L = GM_0^2(-4E_0)^{-1} \, , \quad \mbox{and} \quad T = GM_0^{5/2}(-4E_0)^{-3/2} \, .
\end{equation}
Application of the virial theorem shows that $L$ is the virial radius of the cluster
and $T$ is of the order of the initial dynamical (crossing) time.  In the code,
we use as the unit of time the initial relaxation time, $t_r$, which is given by
\begin{equation}
\label{time2}
t_r = {N_0 \over \ln N_0} T \,  ,
\end{equation}
where $N_0$ is the initial total number of stars in the cluster.  For a
two-component cluster, it is possible to define similar relaxation times
for each component separately.  The relaxation time for component 1,
$t_{r1}$, is given by
\begin{equation}
\label{time3}
t_{r1} = {N_1 \over \ln N_1} T_1 \,  ,
\end{equation}
where $N_1$ is the initial number of stars of component 1, and $T_1$ is the crossing
time for component 1, given by \eq{time1} with $M_0$ and $E_0$ replaced by the 
initial total mass and energy of component 1, respectively.  The relaxation 
time for component 2, $t_{r2}$, is given analogously, although the resulting
expression is only applicable when the two species have decoupled (e.g., 
a collection of black holes undergoing the Spitzer mass-stratification instability).

To facilitate
comparison with previous studies, we report times in this paper
in units of the initial half-mass relaxation time, 
$t_{\rm rh}$, given by the standard expression (see, e.g., Spitzer 1987)
\begin{equation}
\label{time4}
t_{\rm rh} = 0.138 {N_0^{1/2}r_h^{3/2} \over \bar{m}^{1/2}G^{1/2}\ln N_0} \, ,
\end{equation}
where $r_h$ is the initial half-mass radius, and $\bar{m}=M_0/N_0$ is the average
stellar mass.

Each calculation is terminated when the $0.35\%$ Lagrangian radius of the 
heavier component falls below 0.001 (in our length units).
The Lagrangian radii are then inspected graphically and the 
core collapse time, $t_{\rm cc}$, determined by noting the time at which the innermost Lagrangian
radii of the heavier component begin to dip appreciably.  Core collapse times 
determined in this way are not sensitive to the precise criterion used but can 
have a large statistical uncertainty, particularly when the number of stars within 
the innermost Lagrangian radii is small.

\section{Light Tracers}

Intuition has it that globular clusters cannot contain extremely light stars, since
equipartition of energy would imply a mean velocity for the light stars 
that far exceeds the escape velocity for the cluster \citep{taillet95}.  Although proportionately more
light tracers than background stars are lost during the evolution, we find that a significant
fraction are retained in the halo. In some cases their number density there actually increases
during the evolution.

\begin{figure*}[t]
\figurenum{1}
\epsscale{2.1}
\plottwo{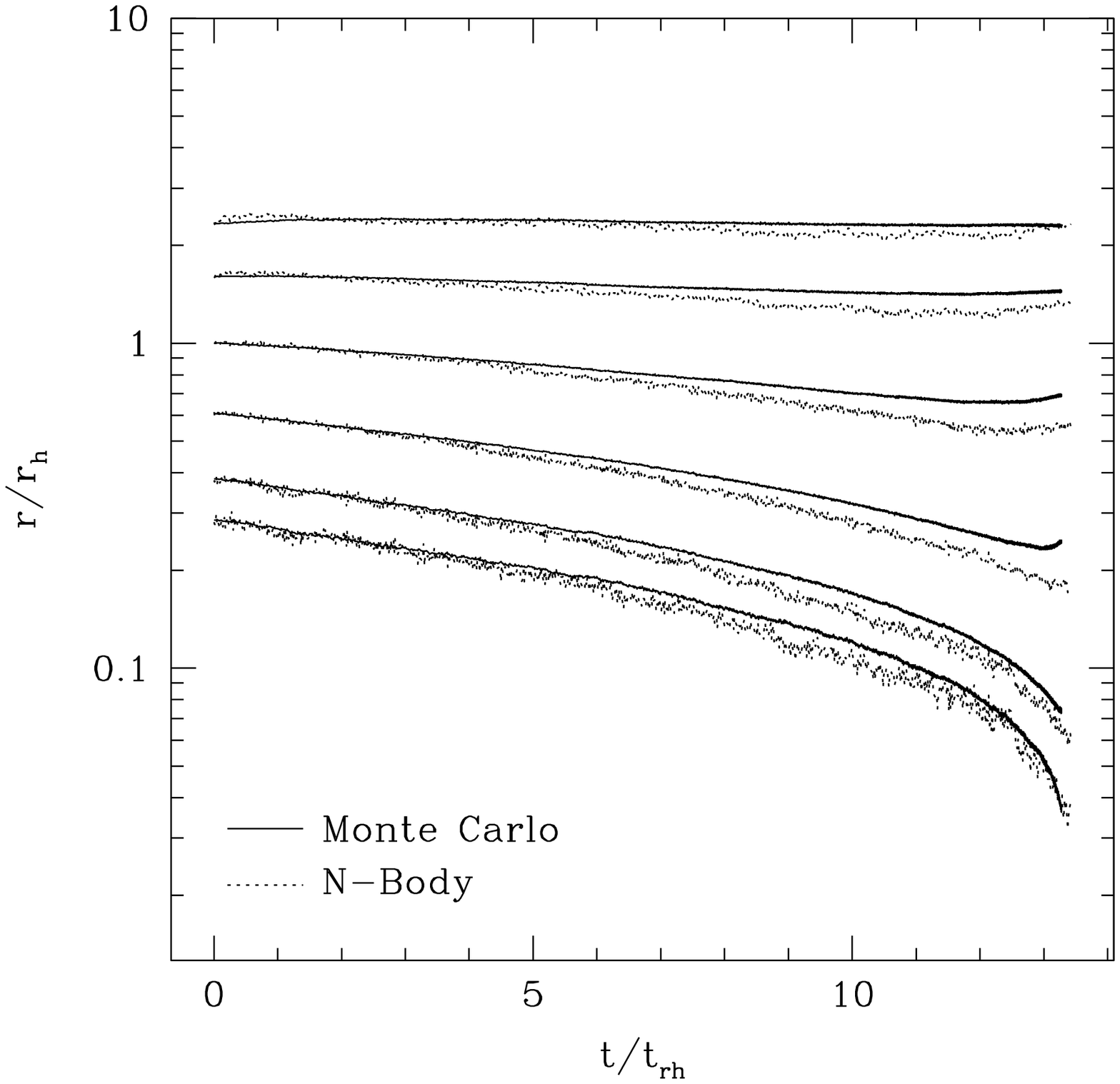}{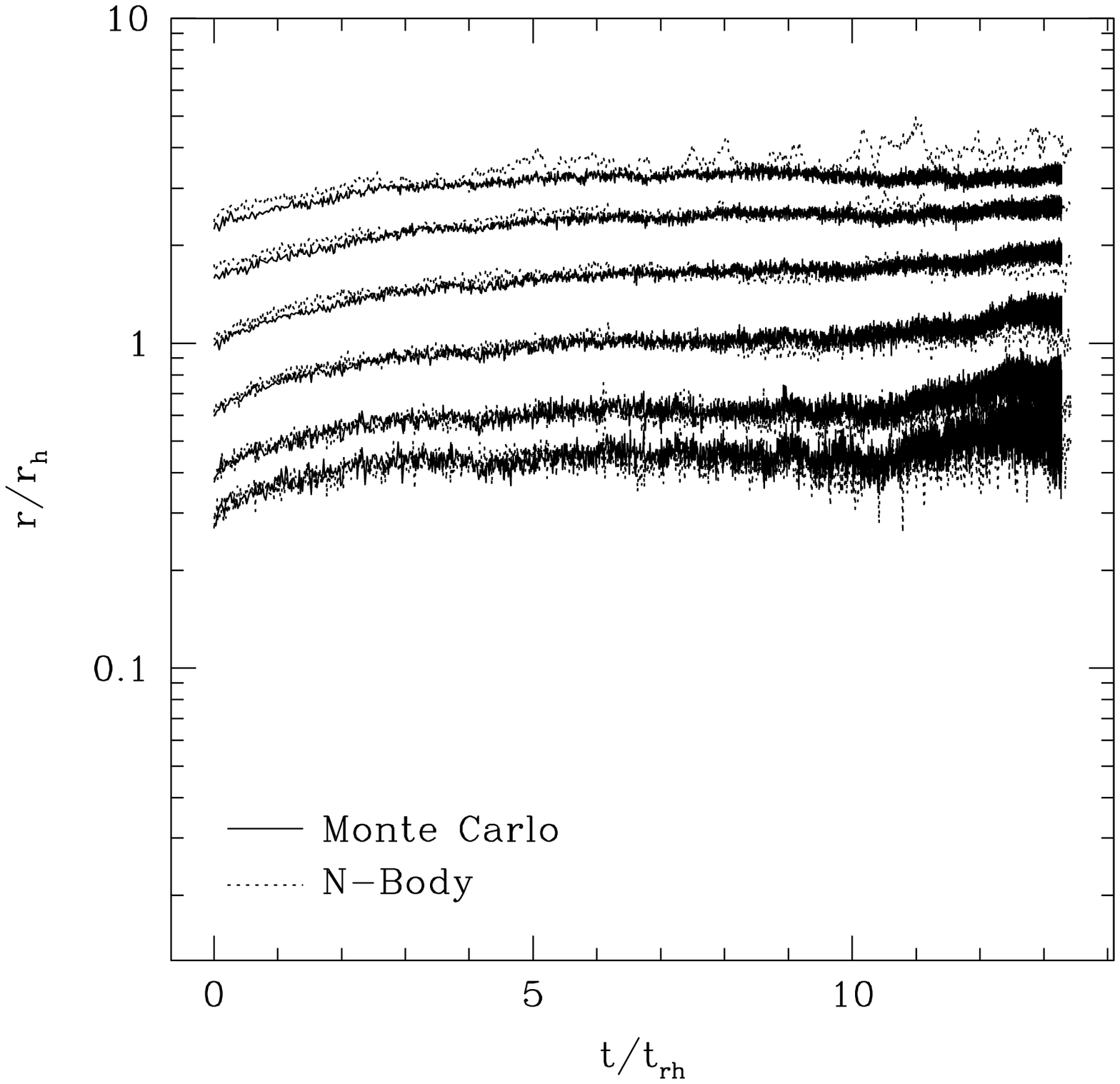}
\epsscale{1}
\caption{\label{fig1}Comparison of Monte Carlo and {\nbody} data.  (Solid line is Monte Carlo, dotted line is 
	{\nbody}.)  Shown are the 5, 10, 25, 50, 75, and 90\% Lagrangian radii of background stars (left plot) 
	and tracers (right plot) for a $W_0 = 5$ King initial model with a mass ratio $\mu = 0.001$, 
	showing reasonable agreement between the two methods.}
\end{figure*}

\begin{figure*}[b]
\begin{minipage}[t]{0.47\linewidth}
\begin{inlinefigure}
\figurenum{2a}
\plotone{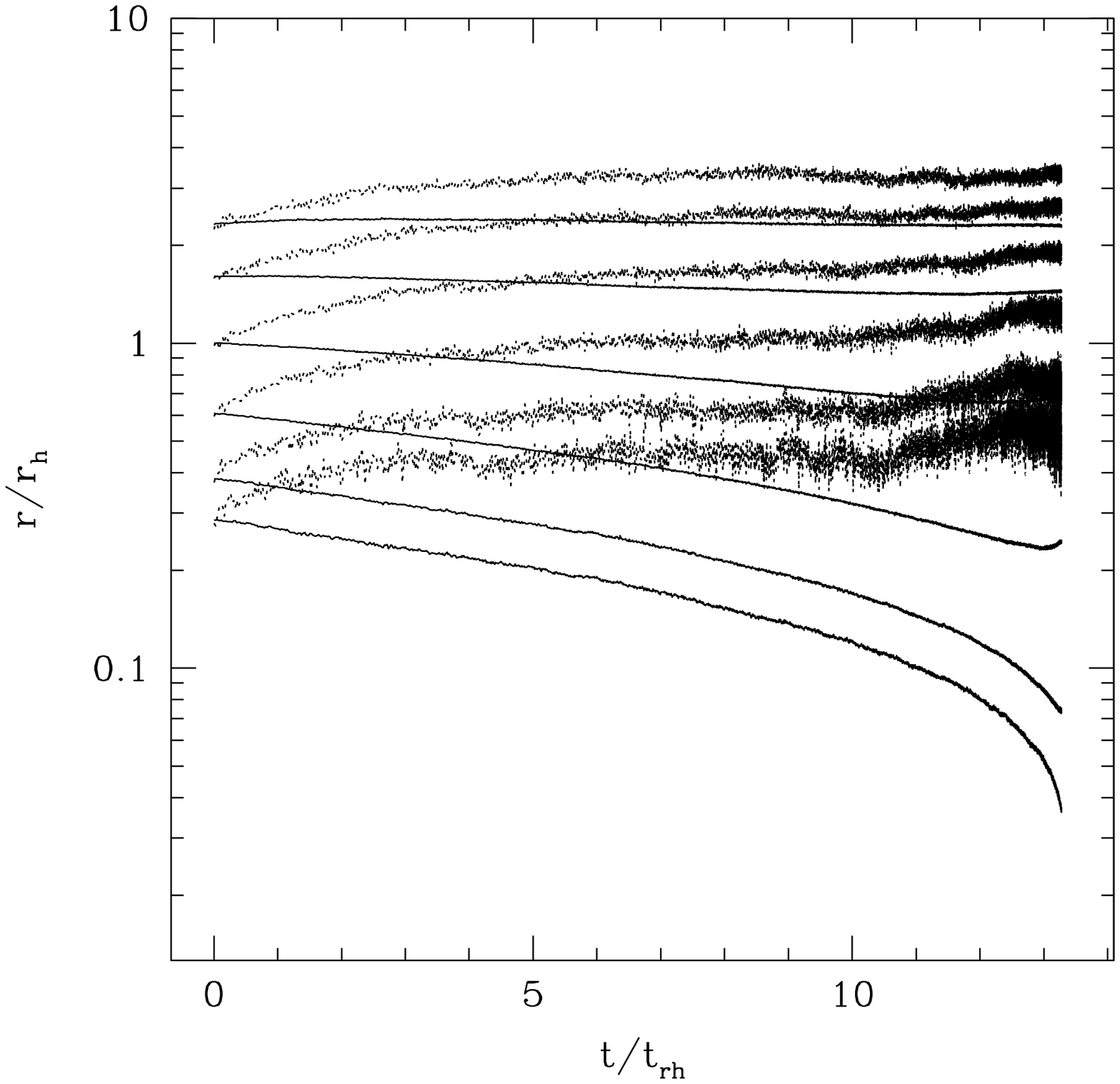}
\caption{\label{fig2a}Lagrangian radii of background stars (solid line) and tracers (dotted line) from \fig{fig1}, 
	showing only the Monte Carlo results.  Note the stark contrast between the two,
	illustrating the extreme mass segregation present even early in the simulation,
	at $t \sim 5 \, t_{\rm rh}$.  By core collapse, over 90\% of the tracers occupy the outer region
	of the cluster in which 50\% of the background stars reside.}
\end{inlinefigure}
\end{minipage}
\hfill
\begin{minipage}[t]{0.47\linewidth}
\begin{inlinefigure}
\figurenum{2b}
\plotone{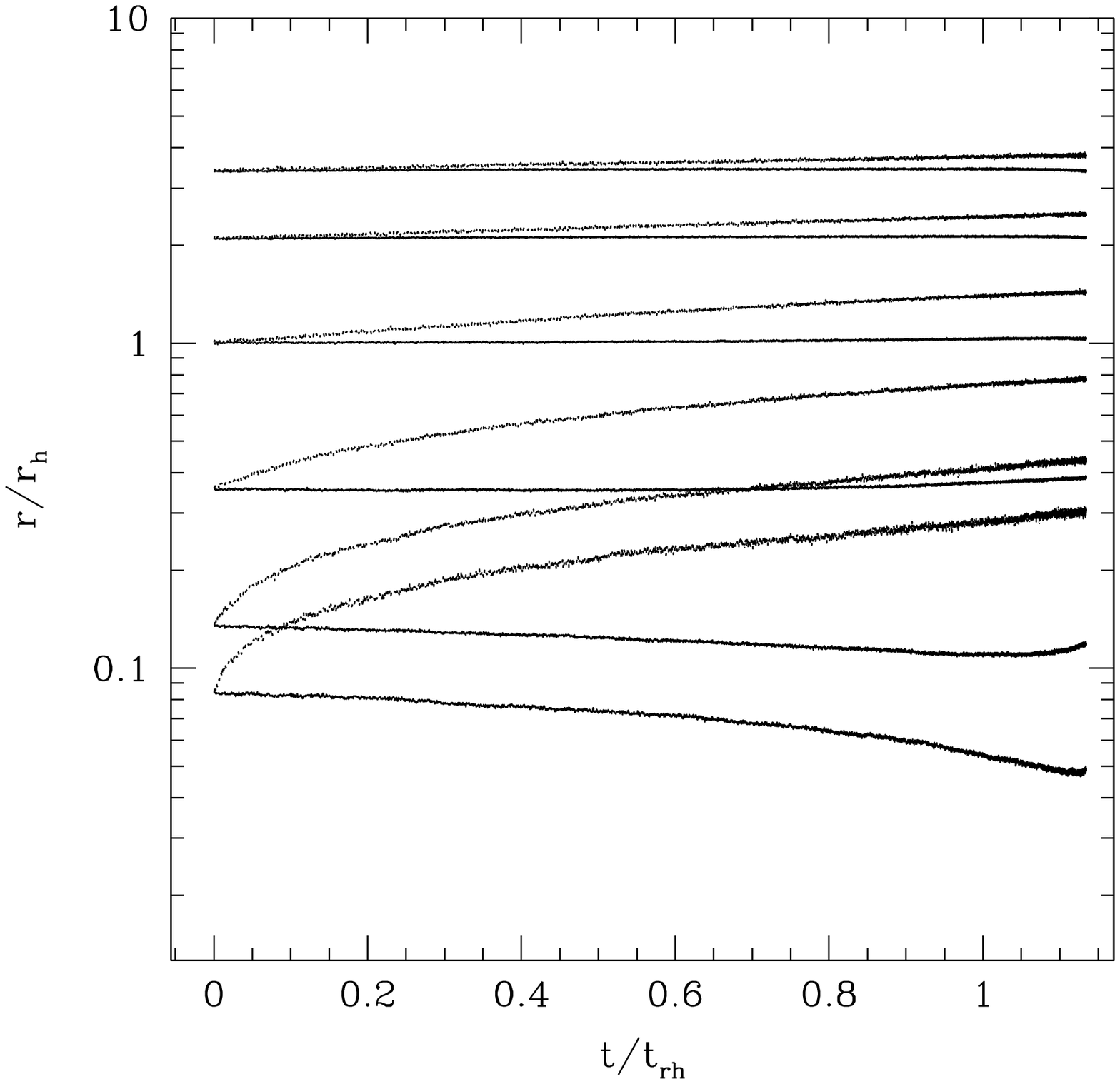}
\caption{\label{fig2b}Same as \fig{fig2a}, but for a very centrally concentrated $W_0 = 9$ 
	King initial model.  Even though this cluster reaches core collapse very quickly (because it is 
	initially almost core-collapsed), by that time over 90\% of the tracers occupy the outer region of 
	the cluster in which 75\% of the background stars reside.}
\end{inlinefigure}
\end{minipage}
\end{figure*}

\subsection{Numerical Results}

\begin{figure*}[t]
\begin{minipage}[t]{0.47\linewidth}
\begin{inlinefigure}
\figurenum{2c}
\plotone{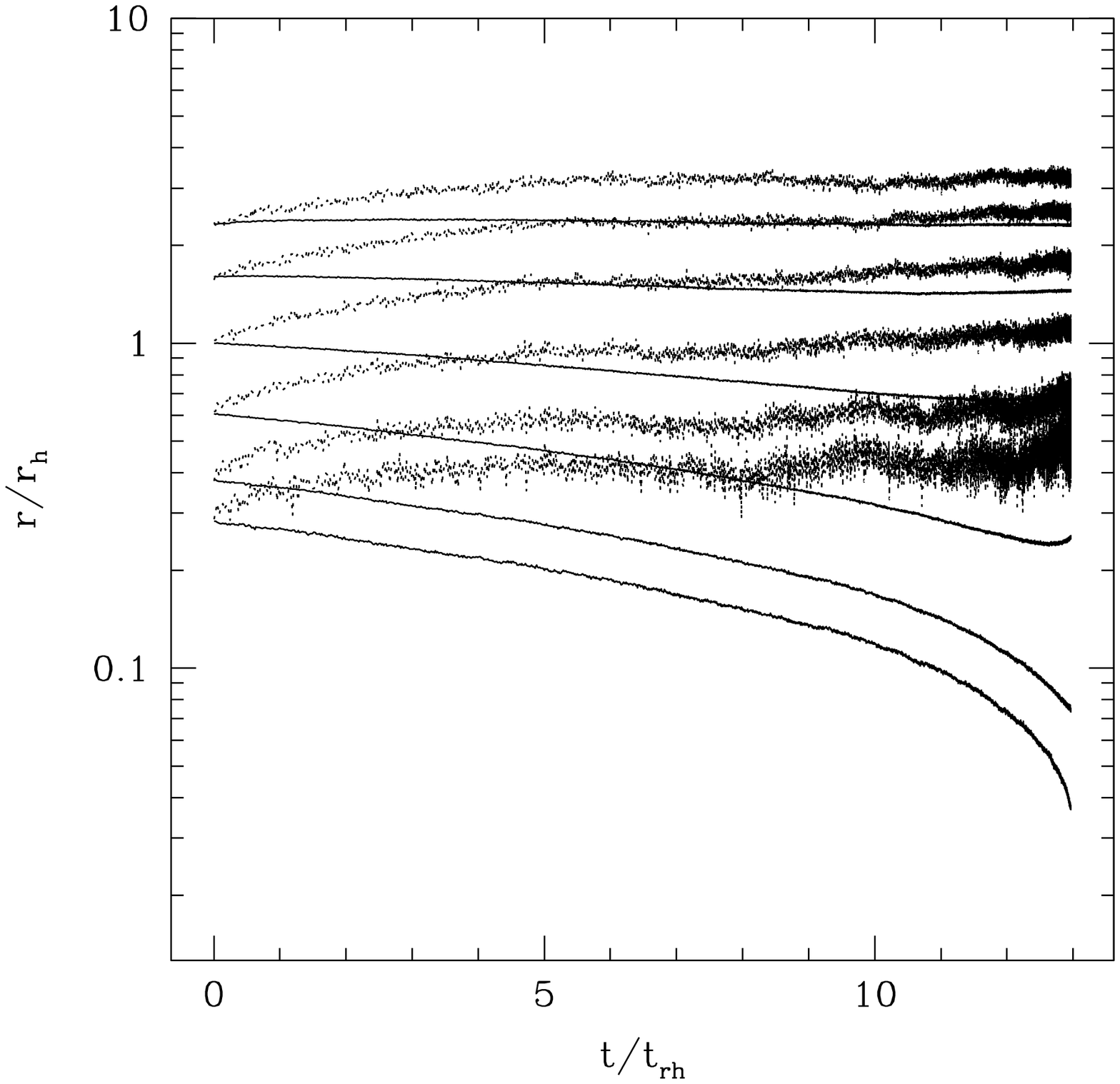}
\caption{\label{fig2c}Same as \fig{fig2a}, but for a mass ratio $\mu = 0.1$.  Almost the same degree of 
	mass segregation present for the extreme mass ratio of $\mu = 0.001$ is witnessed for this more 
	reasonable value.  Again, by core collapse, about 90\% of the tracers occupy the outer region of the 
	cluster in which 50\% of the background stars reside.}
\end{inlinefigure}
\end{minipage}
\hfill
\begin{minipage}[t]{0.47\linewidth}
\begin{inlinefigure}
\figurenum{3a}
\plotone{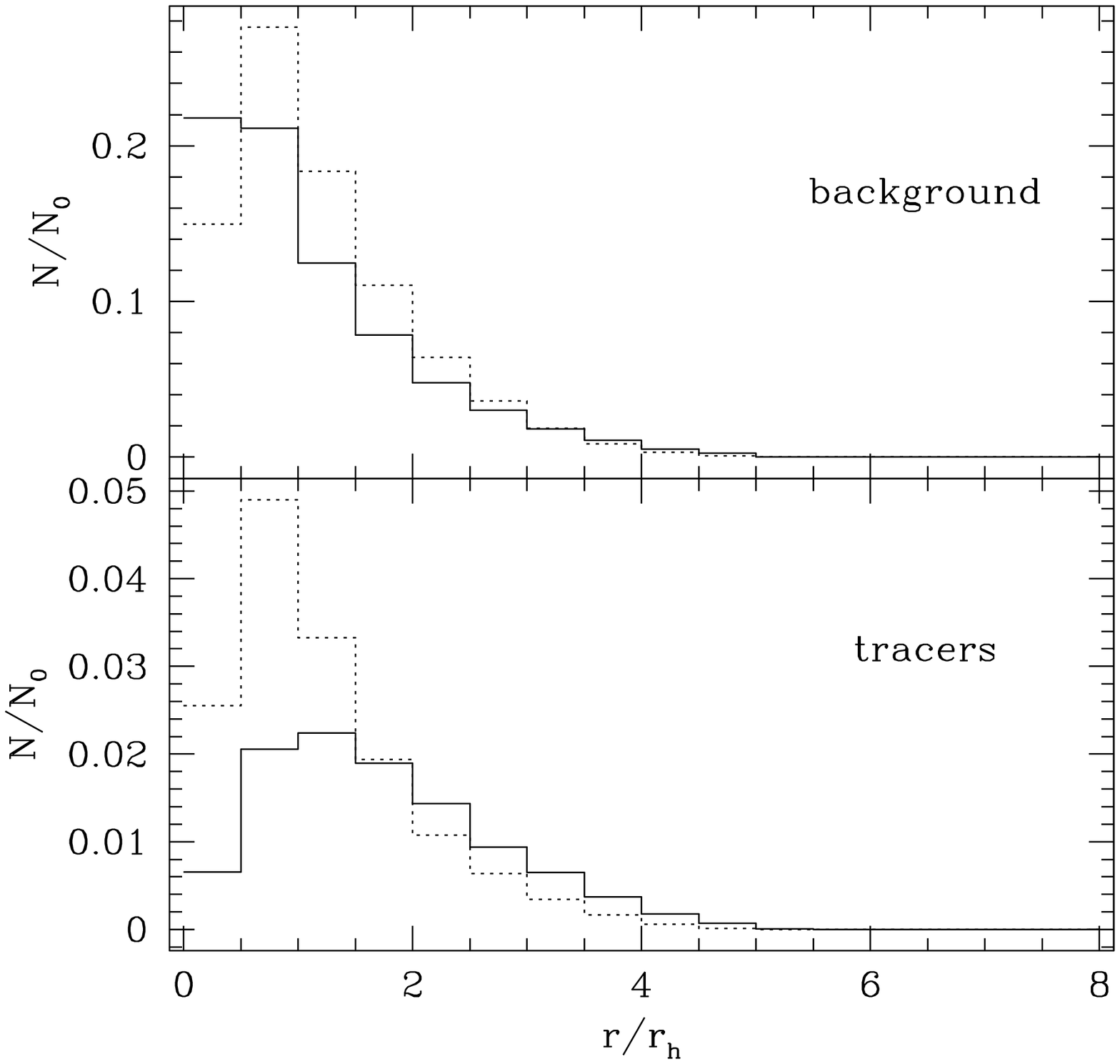}
\caption{\label{fig3a}Number histogram of stars versus radius for a $W_0 = 5$ King initial model
	with a mass ratio $\mu = 0.001$, showing background stars in the top plot
	and tracers in the bottom.  The dotted line represents a time very near
	the start of the simulation, while the solid line represents a time
	midway through the simulation, at $t = 4.9 \, t_{\rm rh}$.  While this {\it isn't}
	a number density plot, it gives a feel for the radial profile of the two species.}
\end{inlinefigure}
\end{minipage}
\end{figure*}

To ensure the validity of our Monte Carlo results, we first sought to compare 
integrations performed with the Monte Carlo code
with the results of direct {\nbody} simulations.  
As a representative case, \fig{fig1} shows 5, 10, 25, 50, 75, and 90\% Lagrange
radii (radii containing a constant mass fraction) 
for background stars (left) and tracers (right) for a $W_0 = 5$ King initial model
with mass ratio $\mu = 0.001$.  The agreement remains quite good at all times. The two 
methods agree remarkably in the most relevant result: that the light tracer stars ``diffuse''
into the outer halo, while the core, comprised mainly of heavier ``background'' stars, contracts.
This is consistent with the mass segregation found with previous {\nbody} and Fokker-Planck
studies \citep{spitzer87}.

\fig{fig2a} displays the segregation more clearly, showing {\it only} the Monte Carlo results from \fig{fig1}.
Here background stars are represented by a solid line, light tracers by a dotted line.
As one expects from simple energy equipartition arguments, the tracers gain energy from interactions
and are pushed out of the core, as is seen in the mass segregation evident as early as
$t \simeq 5 \, t_{\rm rh}$.  However, 25\% of the tracer population is left to linger in the 
outer regions at the time of core collapse.  (By contrast, 65\% of the background stars remain
in the cluster.)  Instead of receiving large monolithic kicks from the heavy stars 
and being immediately ejected from the cluster, the light tracers slowly gain energy and are gradually
pushed out into the halo, where there are so few heavy stars that the light tracers are
unable to exchange energy and are left there to linger.
\fig{fig2b} is the same as \fig{fig2a}, but for a very centrally concentrated
$W_0=9$ King initial model.  Because this cluster is initially so close to core collapse, it does not 
have much time to evolve; yet by collapse over 90\% of the tracers occupy the outer region of the cluster in
which 75\% of the background stars reside.  This extreme segregation is witnessed even for a more
moderate mass ratio of $\mu=0.1$.  \fig{fig2c} shows the same as \fig{fig2a}, but for a $W_0=5$, $\mu=0.1$ 
model.  Here, by the time of core collapse, about 90\% of the tracers occupy the outer region in 
which 50\% of the tracers reside.

\begin{figure*}[t]
\begin{minipage}[t]{0.47\linewidth}
\begin{inlinefigure}
\figurenum{3b}
\plotone{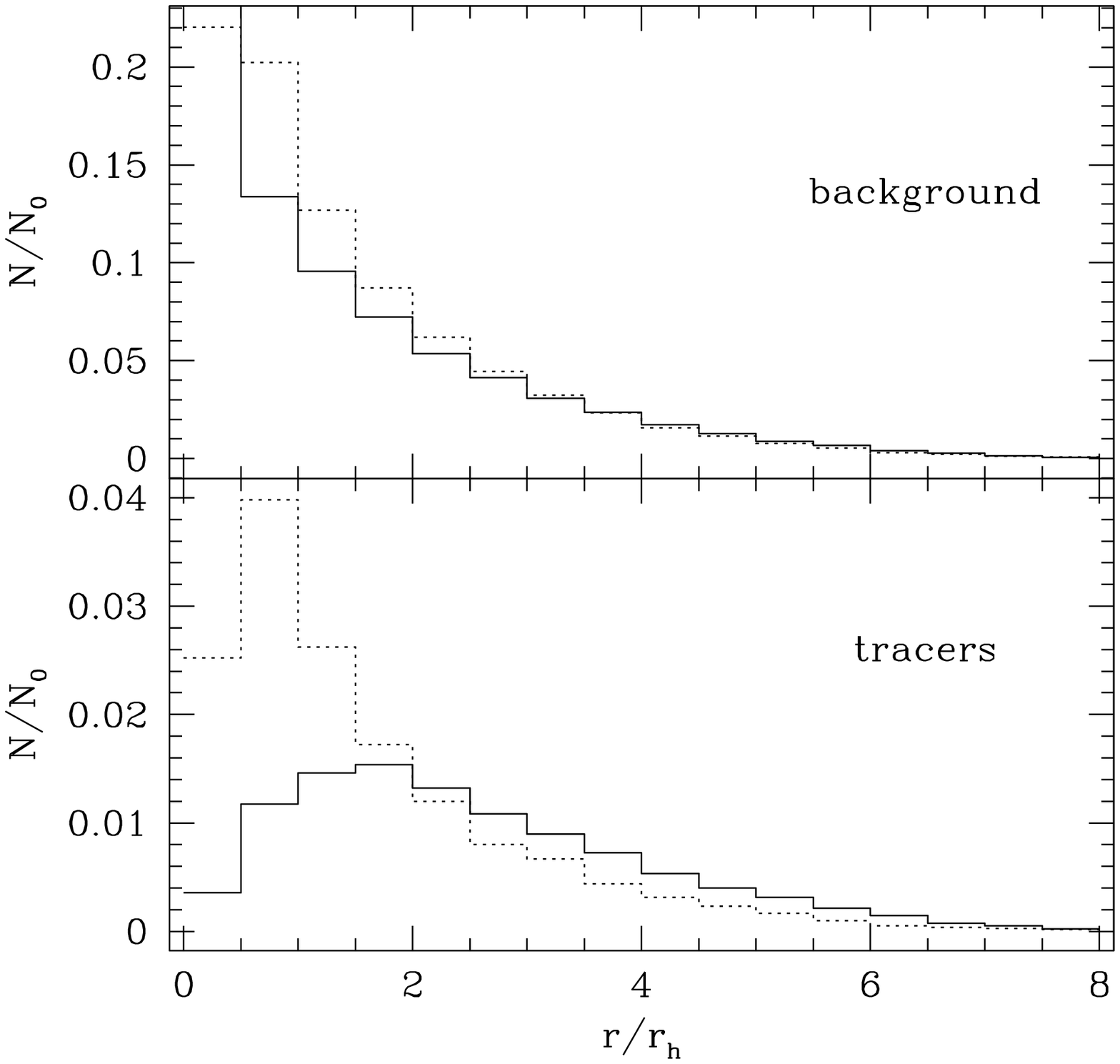}
\caption{\label{fig3b}Same as \fig{fig3a}, but for a $W_0 = 7$ King initial model.  Here 
	the solid line represents a time near collapse, $t = 7.3 \, t_{\rm rh}$.  In this case, the number 
	density of tracers in the halo is very visibly greater at core collapse than at the start of the 
	simulation, and by about 50\% in some regions.}
\end{inlinefigure}
\end{minipage}
\hfill
\begin{minipage}[t]{0.47\linewidth}
\begin{inlinefigure}
\figurenum{3c}
\plotone{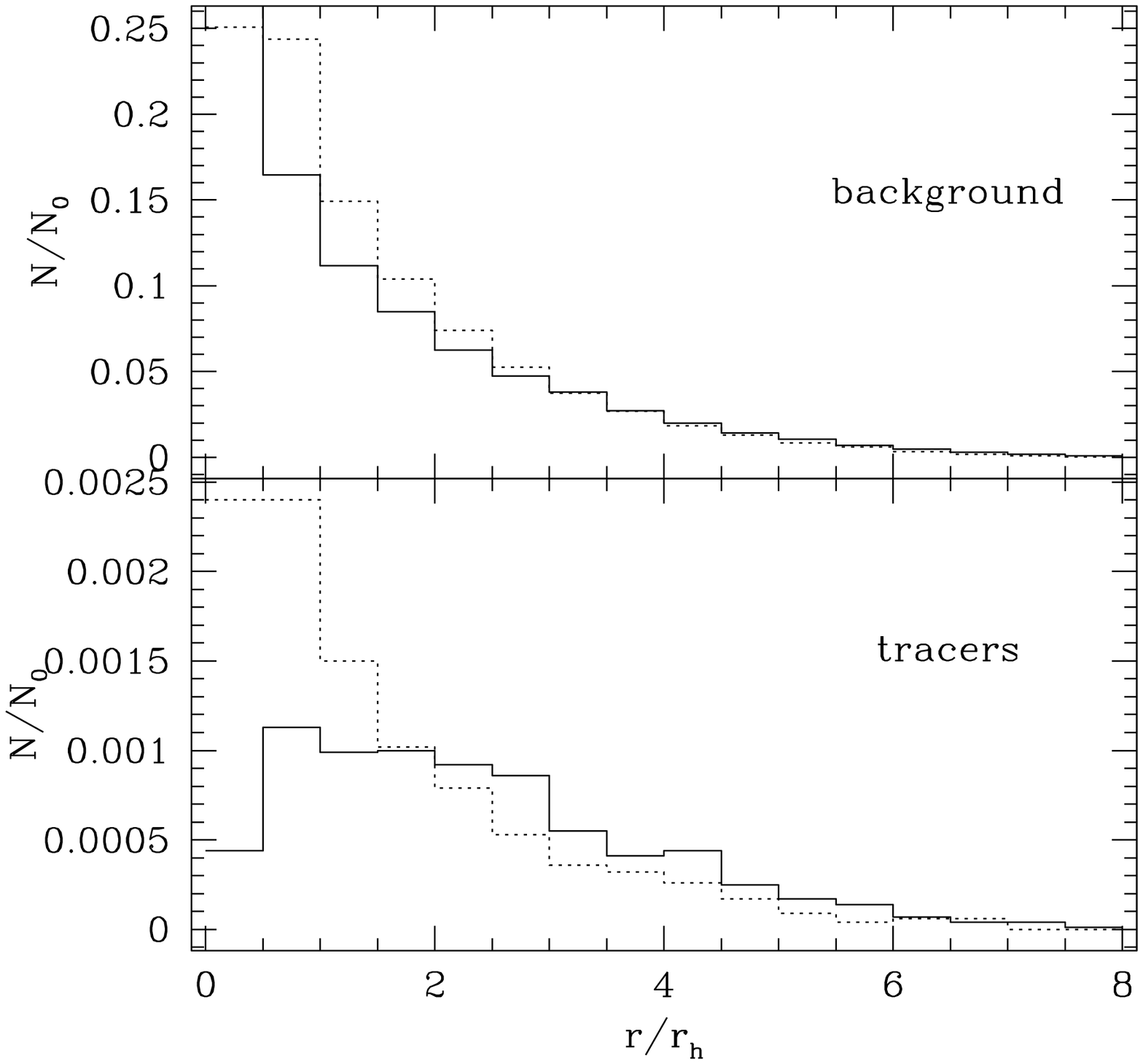}
\caption{\label{fig3c}Same as \fig{fig3a}, but for a $W_0 = 7$ King initial model with a 
	mass ratio $\mu = 0.1$.  Again, the solid line represents a time near core collapse, 
	$t = 7.2 \, t_{\rm rh}$.  Although the curve is a bit noisy, there is clearly an enhancement in 
	the number density of tracers in the halo, even at this modest mass ratio.}
\end{inlinefigure}
\end{minipage}
\end{figure*}

For a more detailed look at the radial profile of the light tracers relative to the background, 
\fig{fig3a} shows the number of stars in each radial bin.  The number in each bin is 
divided by the {\it initial} total number of stars in the cluster, $N_0$, 
so that the vertical axis represents
the fraction of stars in each bin.  This histogram, which is proportional to
$\frac{dN}{dr}$, is {\it not} a number density plot but still gives a feel for the number density,
and was chosen because it most clearly displays phenomena at large radius.
Shown is the same $W_0=5$, $\mu=0.001$ model as in \fig{fig1}, with
background stars in the top plot and tracers in the bottom.  The dotted line represents a time
very near the start of the simulation, while the solid line represents a time midway through
the simulation, at $t=4.9 \, t_{\rm rh}$.  There is a clear increase in the density of background
stars in the core during the evolution, while there is also a clear decrease for tracers.
More importantly, there is a significant {\it enhancement} of tracers in the halo region.
Similarly, \fig{fig3b} shows the same results for a $W_0=7$, $\mu=0.001$ model.  In this case, the 
number density of tracers in the halo is very visibly greater at core collapse than at the 
start of the simulation, and by about 50\% in some regions.  Even for the more modest
mass ratio of $\mu=0.1$, a similar profile is obtained, as shown in \fig{fig3c}.  Although
a bit noisier than the preceding, since only 1000 tracers are used here, the same significant
enhancement of tracers in the halo is evident.

Table 1 gives relevant initial conditions for all the light tracer systems considered here,
as well as core collapse times, the fraction of each species left in the cluster at core collapse, and the ratio
of half-mass radii for the two species at core collapse.  We have also included a few astrophysically
relevant timescales: $\tau_{1.5} (r_c)$, the time for the number of light stars within the initial 
core radius to decrease by a factor of 1.5; $\tau_{10} (r_c)$, the time for the number of light 
stars within the initial core radius to decrease by a factor of 10; $\tau_{1.5} (r_h)$, 
the time for the number of light stars within the initial half-mass radius to decrease by a factor of 1.5; 
and $\tau_{10} (r_h)$, the time for the number of light stars within the initial half-mass radius 
to decrease by a factor of 10.  \fig{fig4} shows a representative determination of these timescales.  
Smooth curves were fit to the data using {\tt gnuplot}'s `acsplines' (approximation cubic splines) 
routine.  The timescales show a clear trend, increasing as the mass ratio approaches unity, consistent
with energy equipartition arguments.  A closer look reveals a critical mass ratio, around 0.3, below 
which the timescales $\tau_{10}$, and especially $\tau_{1.5}$, are roughly constant: the light stars are
so light that they are immediately ejected from the inner regions of the cluster on a timescale that is 
independent of mass
ratio.  A simple derivation of this critical mass ratio, $\mu_{\rm crit}$, proceeds as follows.  The escape
speed from the center of the cluster is given by $v_e^2 = 2 \Psi(0) = 2 W_0 \sigma^2$.  In the 
core of the cluster, densities are very high and it is expected that the two stellar species will attain thermal 
equilibrium: $\langle v_1^2 \rangle_{\rm core} = \mu \langle v_2^2 \rangle_{\rm core}$.  Since the cluster
is dominated by the heavy species (species 1), its velocity dispersion can be obtained directly from the 
distribution function (\eq{one}):
\begin{equation}
\langle v_1^2 \rangle_{\rm core} \equiv 
{\int_0^{\sqrt{2\Psi(0)}} \! v^2 dv f(0,v) v^2 \over \int_0^{\sqrt{2\Psi(0)}} \! v^2 dv f(0,v)} \, .
\end{equation}
This equation can be integrated numerically to give $\langle v_1^2 \rangle_{\rm core}$ in units of
$\sigma^2$: $\langle v_1^2 \rangle_{\rm core} \equiv \alpha \sigma^2$.  Setting the velocity
dispersion of the lighter species equal to the escape speed at the center yields a simple
equation for the critical mass ratio,
\begin{equation}
\label{four}
\mu_{\rm crit} = {\alpha \over 2 W_0} \, ,
\end{equation}
below which the light stars are almost immediately ejected from the core (within a core relaxation time), 
but not necessarily from the cluster.
Table 3 gives $\alpha$ and $\mu_{\rm crit}$ for many values of $W_0$.  Comparison with
Table 1 shows that, indeed, below this mass ratio the light stars are ejected from the central
regions of the cluster on a timescale that is independent of mass ratio; above this mass ratio, the 
segregation timescale increases dramatically.

\begin{figure*}
\centerline{\epsfig{file=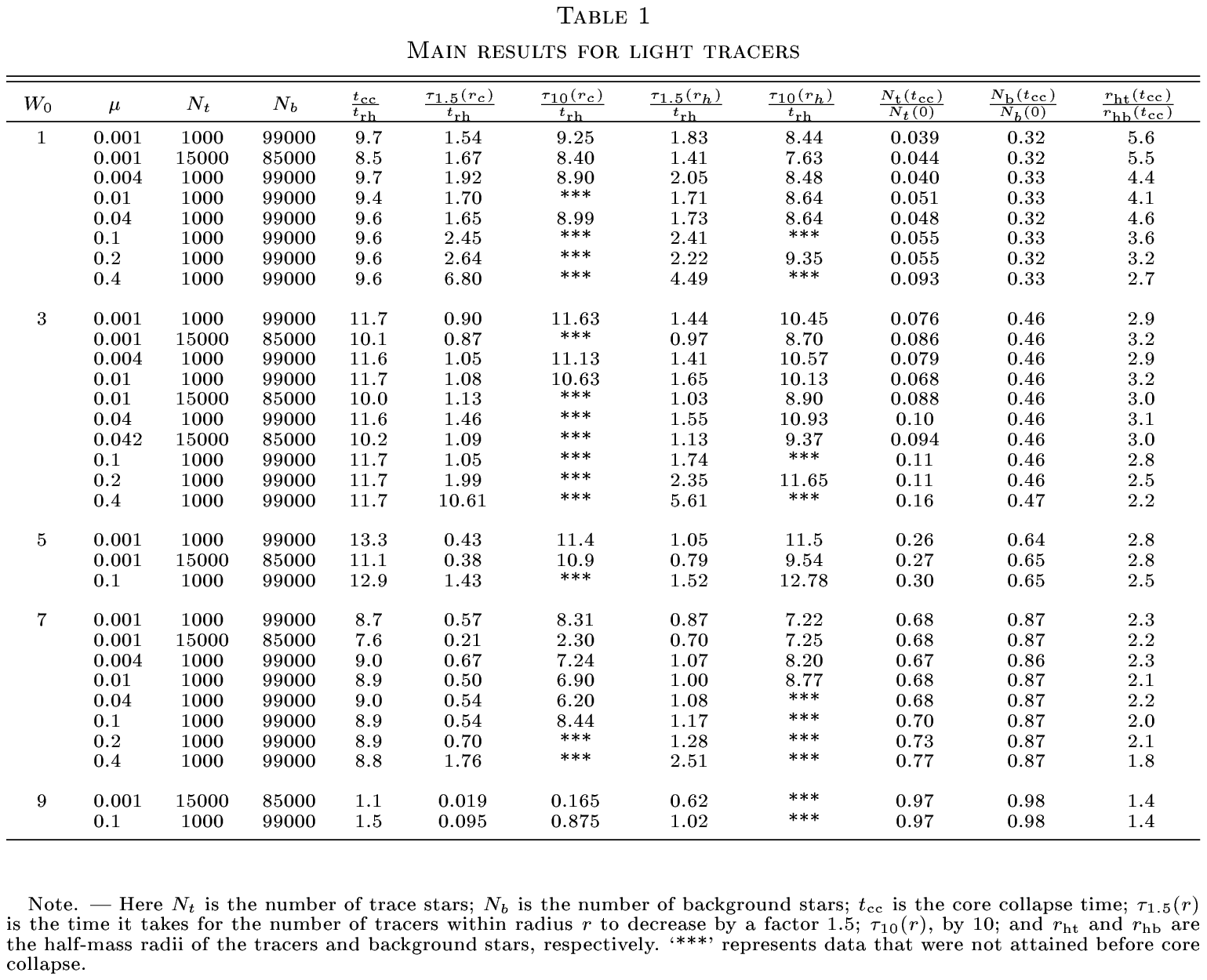}}
\vskip15pt
\centerline{\epsfig{file=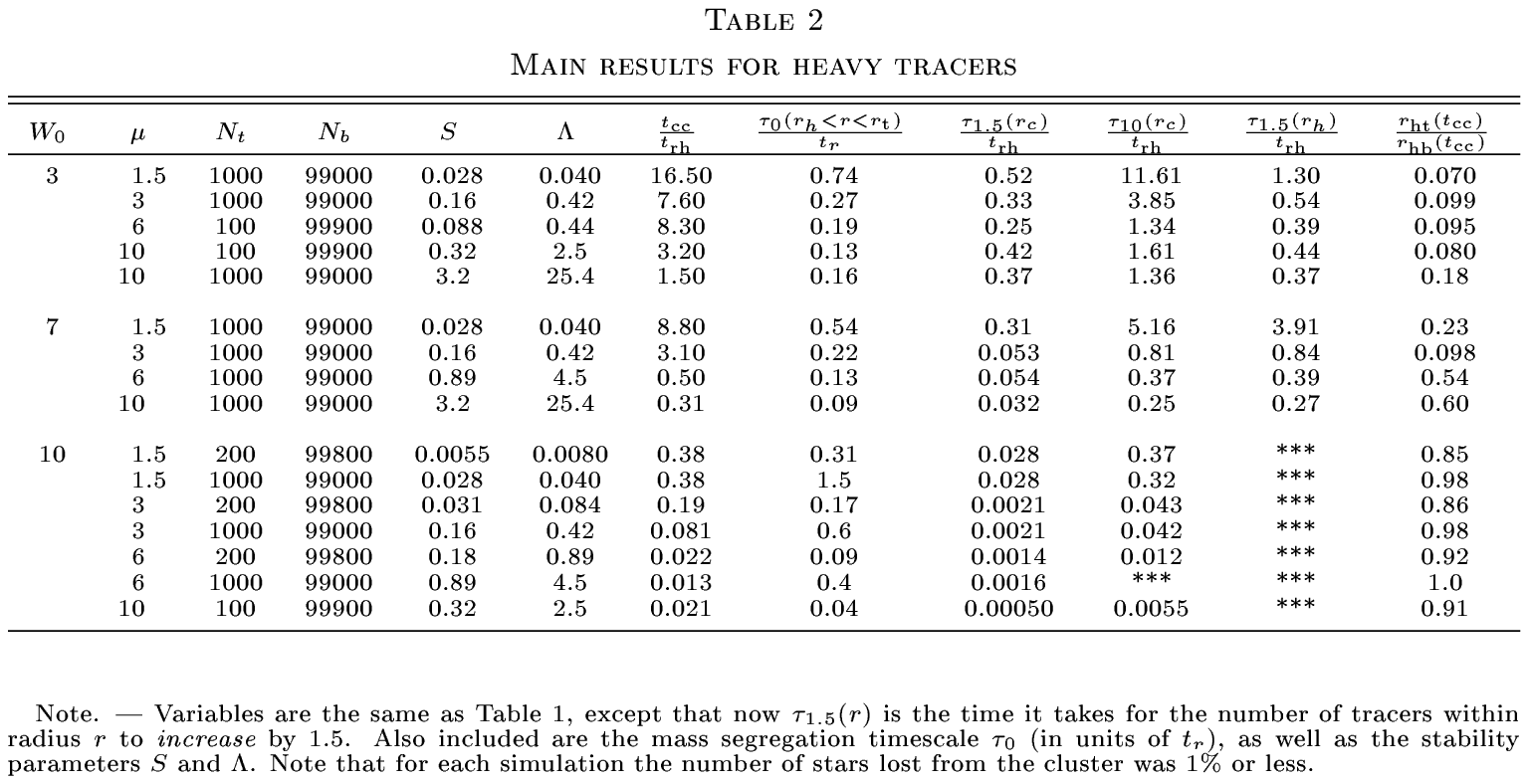}}
\end{figure*}

\begin{figure*}[t]
\begin{minipage}[t]{0.47\linewidth}
\begin{inlinefigure}
\figurenum{4}
\plotone{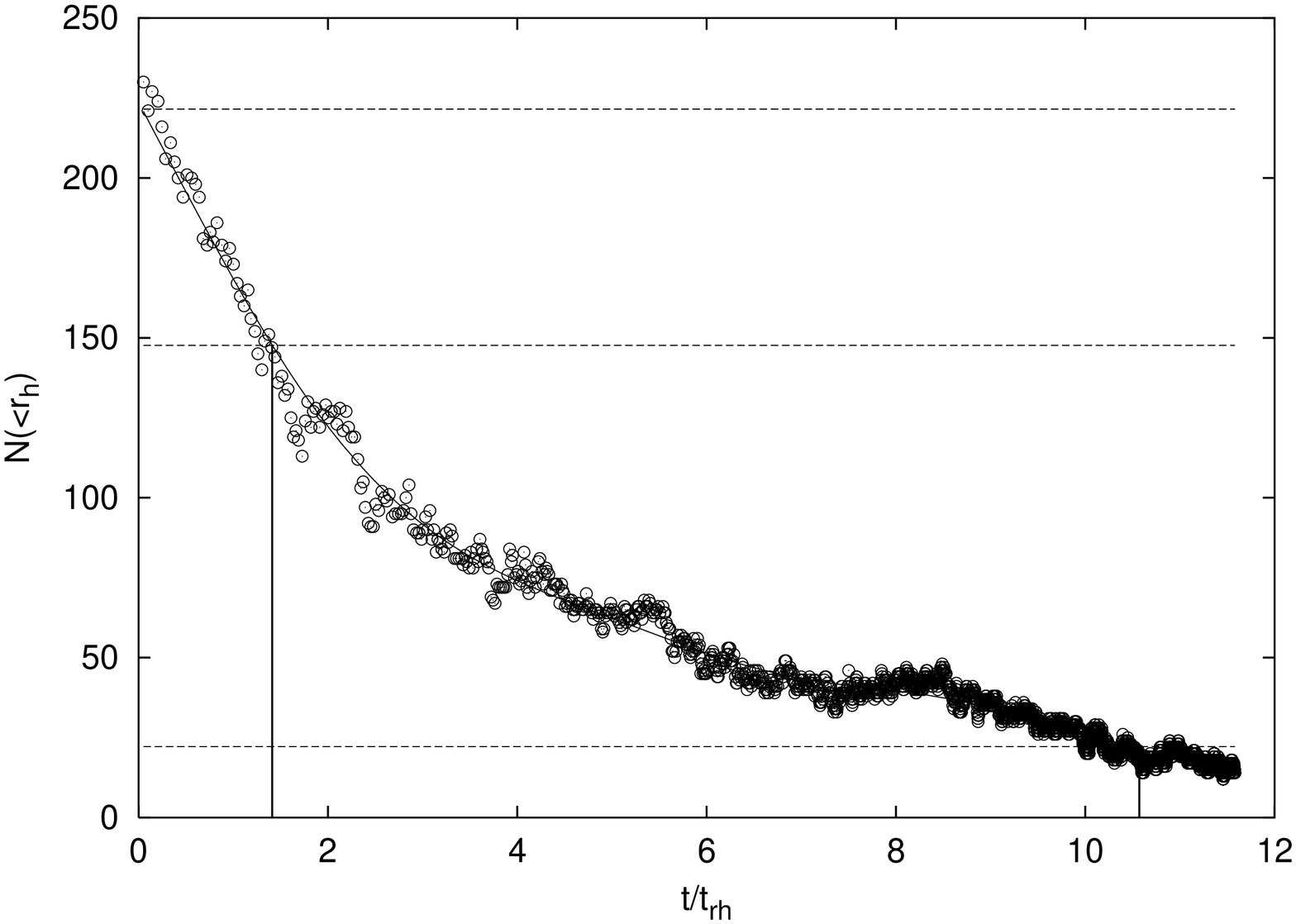}
\caption{\label{fig4}A representative determination of the mass segregation timescales $\tau_{1.5}(r_{\rm h})$
	and $\tau_{10}(r_{\rm h})$ from Table 1 for the case $\mu=0.004$, $W_0 = 3$.  
	The top horizontal line is the number of tracers initially within the half-mass radius, $N_0(r_h)$; 
	the middle horizontal line $N_0(r_h)/1.5$, and the bottom horizontal line $N_0(r_h)/10$.}
\end{inlinefigure}
\end{minipage}
\hfill
\begin{minipage}[t]{0.47\linewidth}
\begin{inlinefigure}
\epsscale{0.82}
\plotone{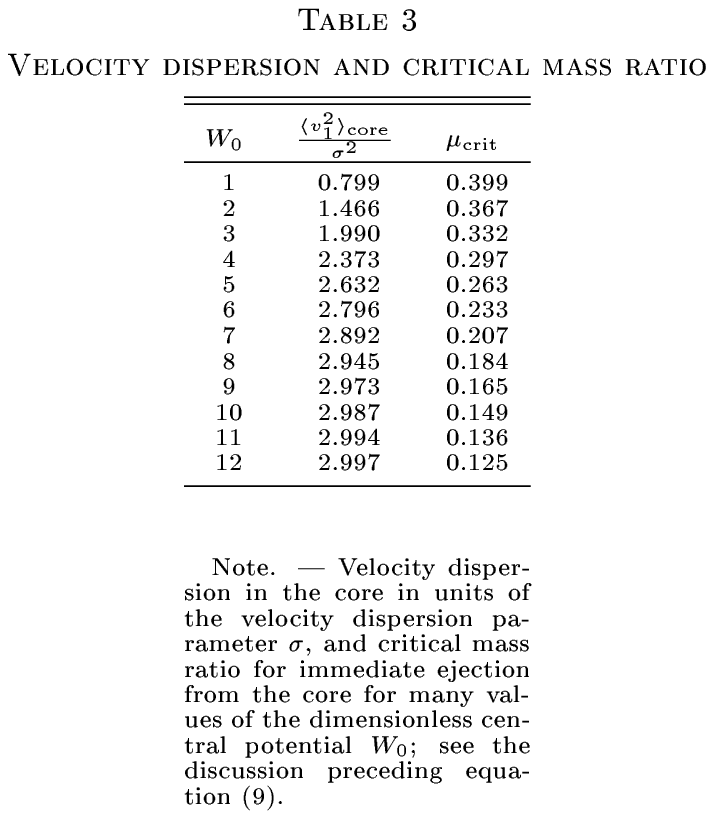}
\epsscale{1}
\end{inlinefigure}
\end{minipage}
\end{figure*}

\subsection{Implications for microlensing}

It is possible that globular clusters could contain copious amounts of 
non-luminous matter---often referred to as (baryonic) dark matter 
in this context \citep{heggie93,taillet95,heggie96}.
However, it is probable that the amount of dark matter is poorly constrained by 
observations of luminous stars.  Consider a cluster with a continuous Salpeter IMF,
\begin{equation}
\label{fourpointfive}
{dN \over dM} \propto M^{-2.35} \, .
\end{equation}
Since we want to study very light objects, let the IMF extend from 0.001 $M_\sun$ to 10 $M_\sun$
(even though such a mass function was never meant to apply below $\sim 0.1 M_\sun$!).
To place this problem in the context of two-component clusters, break up the IMF into two bins: 
a tracer bin extending from 0.001 $M_\sun$ to 0.1 $M_\sun$ corresponding to non-luminous objects,
and a background bin extending from 0.1 $M_\sun$ to 10 $M_\sun$ corresponding to main sequence stars.
(For simplicity, we will ignore stellar evolution and suppose that the heavy stars keep their mass
for the lifetime of the cluster.)  Integrating \eq{fourpointfive} over these two bins gives a mean tracer mass
of 0.0031 $M_\sun$, a mean background mass of 0.31 $M_\sun$, and a ratio of the number of tracers to background 
of $N_t/N_b = 500$.  Using the background stars as species 2 in equations (\ref{two}) and 
(\ref{three}), we find $S=200$ and $\Lambda=1.3 \times 10^4$: this simple model is well outside the Spitzer
stability regime.  The two species will decouple completely and evolve independently; observations of the 
luminous background stars will not betray the presence of the very light tracers.

Approximating the continuous mass spectrum by two bins is most likely unjustified, however, since 
in a continuous mass spectrum stars of mass $M$ will certainly exchange energy with stars of mass $M + dM$.
Thus light stars will exchange energy with slightly heavier stars, which will exchange energy
with slightly heavier stars, etc., providing an indirect means for light objects such as Jupiters to
exchange energy with heavy objects such as main-sequence stars.  Thus thermal equilibrium among all
species may not be
forbidden in a realistic cluster, but the time scale over which it occurs is unclear.

\citet{taillet95} and \citet{taillet96} considered multi-component King models and assumed that thermal equilibrium
{\it did} occur among species.  Using analytical methods, they found that the cluster surface brightness was nearly unaffected
by the amount of non-luminous stars present, for a cluster with dark matter content ranging from 0\% to
90\% by mass.  Therefore, observations of the luminous cluster stars would give no evidence 
of the dark matter in the cluster.  (However, in the most extreme cases, the projected velocity 
dispersion profiles could be used to detect dark matter by observing the luminous component.)

Regardless of whether thermal equilibrium occurs in clusters, it is probable that dark matter would not be easily detectable
by observing the luminous members of clusters.  Dark matter could be directly detected, however, by its
gravitational lensing effect on distant, 
luminous sources \citep{paczynski86,paczynski91,paczynski94}.  By convention, 
a microlensing event is said to occur when a background source lies, in projection, within
the Einstein radius of a lens and is consequently amplified by a factor greater than 1.34 \citep{refsdal,vietri}.
The optical depth, $\tau$, is the probability that such an event will occur, and for low probability,
is given by
\begin{equation}
\label{five}
\tau = {1 \over \delta \omega} \int \! dV n(D_d) \pi \theta_E^2 \, ,
\end{equation}
where $\delta \omega$ is the solid angle of sky considered, $dV=\delta\omega D_d^2 dD_d$ is 
the element of volume, $n$ is the number density of 
lenses, $D_d$ is the distance to the lens (deflector), and $\theta_E$ is the
angular Einstein radius of the lens.  The Einstein radius is given by the usual expression
\begin{equation}
\theta_E^2 = {4 G M \over c^2} {D_s - D_d\over D_d D_s} \, ,
\end{equation}
where $D_s$ is the distance to the source.
Here we will be considering lensing of sources in the Galactic bulge, for example, by globular clusters
along the line of sight to the bulge; therefore, typical values for $D_d$ and $D_s$ are 
$\sim 3$ kpc and $\sim 8$ kpc, respectively.  Since tidal radii of globulars
are typically of order tens of parsecs, we can approximate $D_d$ and $D_s$ as constant over
the range of integration to arrive at the simpler expression
\begin{equation}
\tau = {4 \pi G \over c^2} \Sigma(r) D_d \left(1-{D_d \over D_s}\right) \, ,
\end{equation}
where $\Sigma(r)$ is the surface density along the line of sight, and $r$ is the distance perpendicular
to the line of sight.

\begin{figure*}[t]
\begin{minipage}[t]{0.47\linewidth}
\begin{inlinefigure}
\figurenum{5a}
\plotone{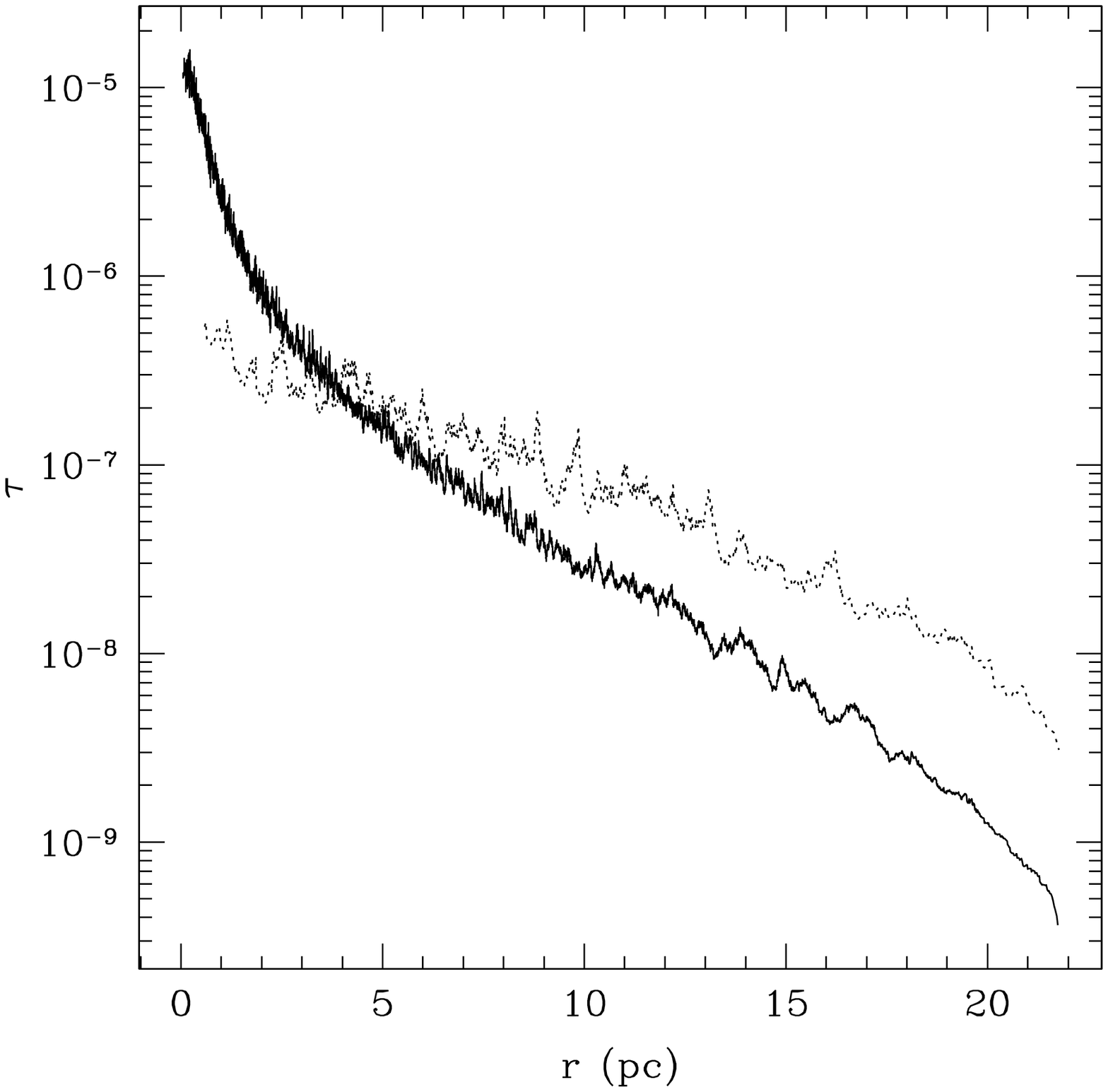}
\caption{\label{fig5a}Optical depth to microlensing due to heavy stars (solid line) and light stars (dotted line)
	for a $W_0=3$ initial King model with 15000 tracers of mass ratio $\mu=0.01$.  The number
	density of light stars has been amplified to emulate a continuous Salpeter IMF.  This snapshot was taken
	near core collapse, at $t = 9.5 \, t_{\rm rh}$.  Generic values of $D_d=3 \, \mbox{kpc}$ and 
	$D_s=8 \, \mbox{kpc}$ were used in the calculation.}
\end{inlinefigure}
\end{minipage}
\hfill
\begin{minipage}[t]{0.47\linewidth}
\begin{inlinefigure}
\figurenum{5b}
\plotone{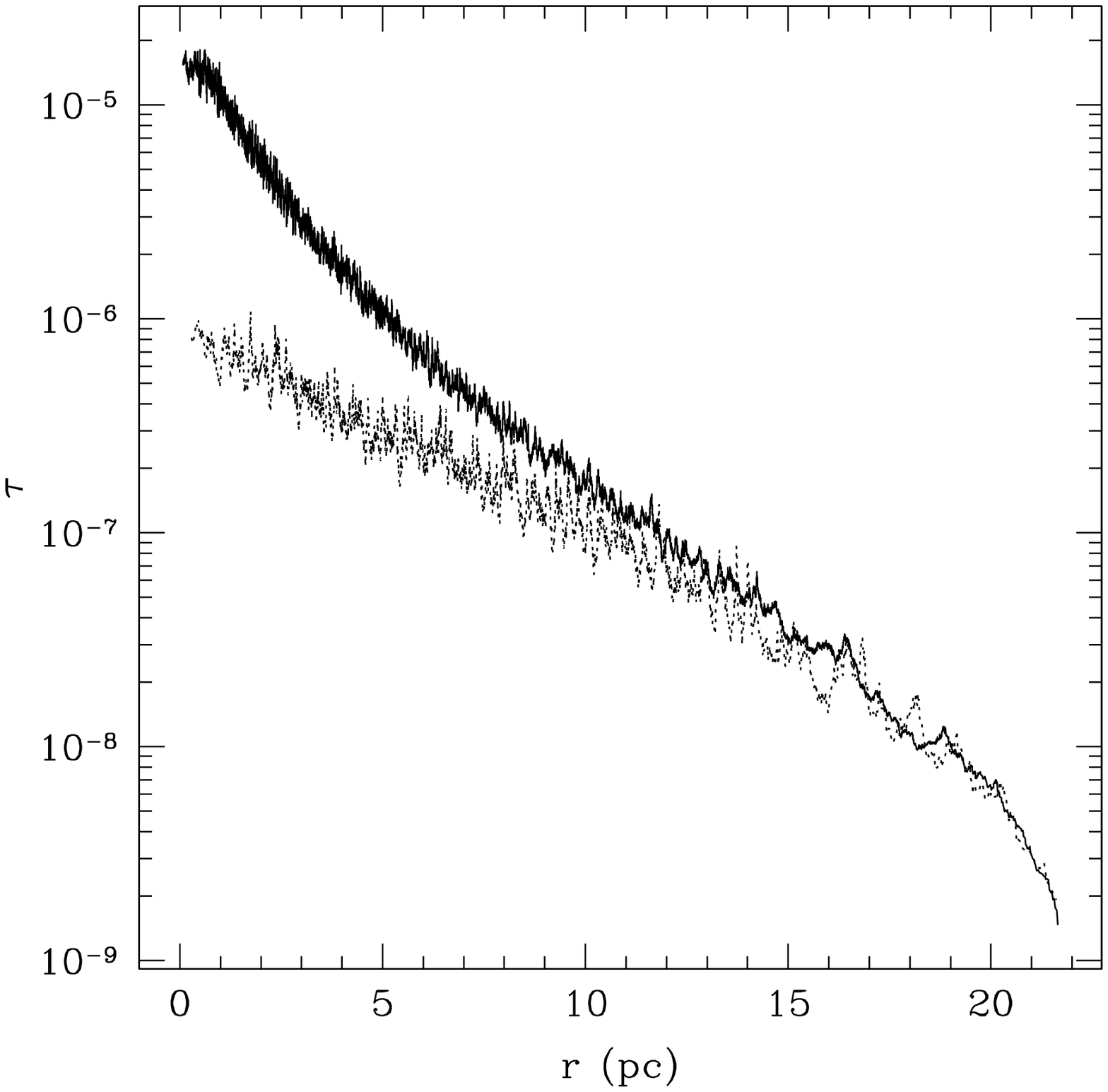}
\caption{\label{fig5b}Optical depth to microlensing at present due to luminous stars (solid line) 
	and dark matter (dotted line) for the globular cluster M22.}
\end{inlinefigure}
\end{minipage}
\end{figure*}

\fig{fig5a} shows the optical depth due to heavy stars (solid line) and light stars (dotted line) for 
a $W_0=3$ initial King model with 15000 tracers of mass ratio $\mu=0.01$.  We have assigned the
background stars a mass of 0.31 $M_\sun$, and multiplied the tracer number density by a factor
$500\times (85000/15000)$, consistent with the preceding discussion of approximating the Salpeter IMF by two mass
bins.  (This heuristic model therefore defines a thermally equilibrized two-component idealization of the 
continuous mass function.)  We have also multiplied both optical depths by a factor of 5
so that the figure is normalized to a cluster that initially had $5 \times 10^5$ stars 
and a mass of $1.3\times 10^5 M_\sun$.  Generic values of $D_d=3 \, \mbox{kpc}$ and $D_s=8 \, \mbox{kpc}$ 
were used in the calculation, and this snapshot of the system was taken at $t = 9.5 \, t_{\rm rh}$.  Note
not only the mass segregation, but also that at large radii the optical depth due to the light stars is 
about an order of magnitude greater than that due to the heavy stars.  Significant numbers of 
non-luminous stars may linger in the outer regions of globular clusters.

\subsection{Dark matter in M22}
\citet{sahu} have carried out {\it HST} observations of the central regions of the globular cluster
M22 (NGC 6656) and have detected at least one microlensing event.  All the parameters needed to deduce
the mass of the lens from the timescale of the lensing event are fairly well known so that
a prediction of its mass can be made.  Assuming distances of 8.2 kpc to the source (for a typical 
star in the Galactic bulge) and 2.6 kpc to M22, as well as a proper motion for M22 of 
10.9 mas yr${}^{-1}$ (134.4 km s${}^{-1}$), they determine the
mass of the lens to be $0.13^{+0.03}_{-0.02} M_\sun$.  This result is exciting because it is the first such detection
where the mass of the lens can be estimated.  They also report a tentative detection
of 6 sub-Jupiter mass objects in the core (an upper limit of $\sim 0.25$ times the mass 
of Jupiter for each) that has since proven spurious \citep{sahu2}.  
Although the events are no longer associated with planets, we still present our theoretical analysis
of the amount of dark matter in M22 as a useful exercise, illustrating how future detections could be used
to estimate the amount of dark matter in clusters.

It is a simple matter to calculate the optical depth due to dark matter given these detections.
Since the lenses are moving with respect to the source, the actual probability for
microlensing is modified from \eq{five} to 
\begin{equation}
\label{six}
{\rm Prob} \equiv {N_e \over N_o}= {1 \over \delta \omega} \int \! dV
n(D_d) 2 \theta_E \dot{\varphi} \Delta t \, ,
\end{equation}
where $N_e$ is the number of microlensing events observed, $N_o$ is the number of observations,
$\dot{\varphi}$ is the angular velocity of the lens, and $\Delta t$ is the time duration of each
observation.  In terms of the optical depth, this is simply
\begin{equation}
{N_e \over N_o} = {2 \over \pi} \tau {\Delta t \over t_0} \, ,
\end{equation}
where $t_0 \equiv \theta_E / \dot{\varphi}$ is the characteristic 
timescale of a microlensing event.  The optical depth is then
\begin{equation}
\label{seven}
\tau = {\pi \over 2} {N_e \over N_o} {t_0 \over \Delta t} \, .
\end{equation}
\citet{sahu} monitored approximately 83,000 stars for a period of $\sim 105$ days, and found
6 candidate dark matter (below 0.1 $M_\sun$) microlensing events lasting at most 0.8 days each.  
\Eq{seven} then gives
an optical depth due to dark matter of $8.7 \times 10^{-7}$.  Adopting the same simple prescription
used before of approximating a continuous mass function by two mass bins, we can evolve a two-component
cluster from a reasonable initial state until its profile resembles that of M22.  We can then
adjust the number of tracers so that the optical depth in the core due to dark matter matches observations,
and consequently estimate what fraction of the cluster mass would have been in dark matter had
the 6 tentative detections proven correct.

For our initial conditions we take a $W_0 = 3$ King model with 85000 background stars of mass
0.38 $M_\sun$ and 15000 tracers of mass 0.016 $M_\sun$ to emulate a Kroupa IMF---which we have
extended well below the $0.1 M_\sun$ lower limit for which it was intended---from 
0.001 $M_\sun$ to 10 $M_\sun$ split into two bins at 0.1 $M_\sun$ \citep{kroupa}.  The Kroupa
IMF is more conservative than a Salpeter in that it contains far fewer sub-stellar-mass 
objects.  (The exact form
of the IMF chosen is largely irrelevant, since in the end we scale the number of stars in each
mass bin to match observations; however, it may affect the percentage of light stars lost
from the cluster during evolution.)  We then
evolve the cluster until its profile matches that of M22, approximately a $W_0=6$ model
\citep{harris}.  We adjust the number of background stars so that the mass due to
luminous matter is $10^{5.57} M_\sun$ \citep{mandushev}, and adjust the number of tracer 
stars so that the optical depth due to dark matter in the core is approximately
$8.7 \times 10^{-7}$.

\fig{fig5b} shows the optical depth due to luminous stars (solid line) and dark matter (dotted line) 
using the prescription just given.  Adopting the total cluster mass and dark matter optical depth quoted
above, we find that the current mass of M22 would have been $\sim 20\%$ dark matter by mass.  
If either the total cluster
mass or optical depth are off by a factor of two, this number could be as small as 12\% or as high
as 35\%.  If both are off by a factor of two, then this number
could be as small as 6\% or as high as 52\%.  This range of values is consistent with the value quoted
by \citet{sahu} of $\sim 10\%$, but implies a higher median.  Using the median values, this analysis 
implies that M22 would have had an initial mass of $8.3 \times 10^5 M_\sun$, 
35\% of which was dark matter (this figure could be anywhere between 12\% and 69\%), and that it 
would have lost 
$2 \times 10^5 M_\sun$ of dark matter to the Galaxy.  These ranges of values should be treated with 
caution, however, since our Monte-Carlo code ignores the effects of binaries, mass loss
due to stellar evolution, and the effects of a non-spherical tidal field, all of which
may affect the fraction of light tracers retained in a cluster.  Furthermore, it is clear
that in order to tighten constraints on the amount of dark matter, one must also refine estimates
of the distances to the lens and the source.

\subsection{Upper Limit on Dark Matter in M22}

All six ``planetary'' events are no longer interpreted as being due to microlensing.
More careful examination reveals that the apparent brightening observed in 
both of a pair of cosmic-ray-split images was actually due to point-like cosmic 
rays hitting close to the same star in both images \citep{sahu2}.  This 
null detection, while unfortunate, can be used to estimate an upper limit
on the mass fraction of M22 in very low-mass objects.

Adopting our simple prescription of approximating the continuous mass function
by two bins, one containing luminous matter (objects heavier than $0.1\, M_\sun$),
the other containing dark matter (objects lighter than $0.1\, M_\sun$), we assume
that the dark matter in M22 is represented by a collection of objects with the average
mass in that bin, $0.016\, M_\sun$.  We then expect the number of dark matter
lensing events in an observation to obey a Poisson distribution.  Adopting a mean of 1 dark matter
lensing event, using $t_0 \approx 6\,{\rm days}$ for a $0.016\, M_\sun$ lens, and
performing essentially the same analysis as in Sec.\ 3.3, a null detection
yields an upper limit on the current mass fraction of M22 in very low-mass 
objects of $\sim 25\%$ at the 63\% confidence level.  Adopting a mean of 3 
events yields an upper limit of $\sim 50\%$ at the 95\% confidence level.

Although our simple calculation is physically motivated, it is necessarily crude, 
since a careful analysis should take into account the mass distribution of the 
dark matter.  In addition, there is another effect to consider.  The typical separation between 
exposures in the \citet{sahu} observation is about 3 days.  Thus a microlensing event
with characteristic width smaller than 3 days, corresponding to a lens mass of
$0.004\,M_\sun$, has a non-zero probability of not being 
detected, biasing detections toward higher masses.  Since the smallest mass we consider is 
$0.001\,M_\sun$, presumably this effect is small.

\begin{figure*}[t]
\begin{minipage}[t]{0.47\linewidth}
\begin{inlinefigure}
\figurenum{6}
\plotone{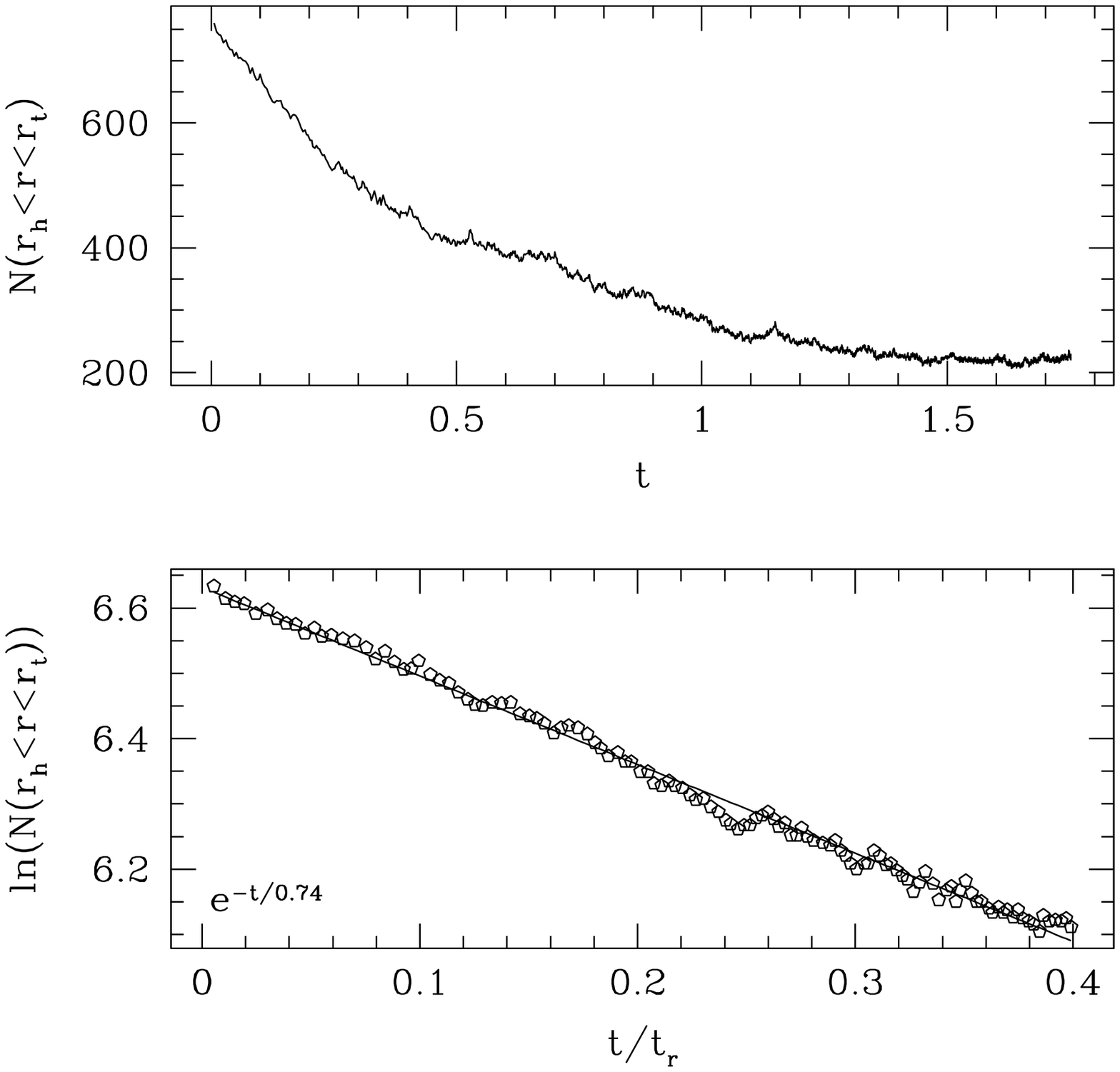}
\caption{\label{fig6}Number of tracers in the halo ($r_h < r < r_t$) as the cluster contracts, showing
	a representative determination of the mass-segregation timescale $\tau_0$ for the case
	$W_0=3$, $\mu=1.5$, and $N_t=1000$.}
\end{inlinefigure}
\end{minipage}
\hfill
\begin{minipage}[t]{0.47\linewidth}
\begin{inlinefigure}
\figurenum{7}
\plotone{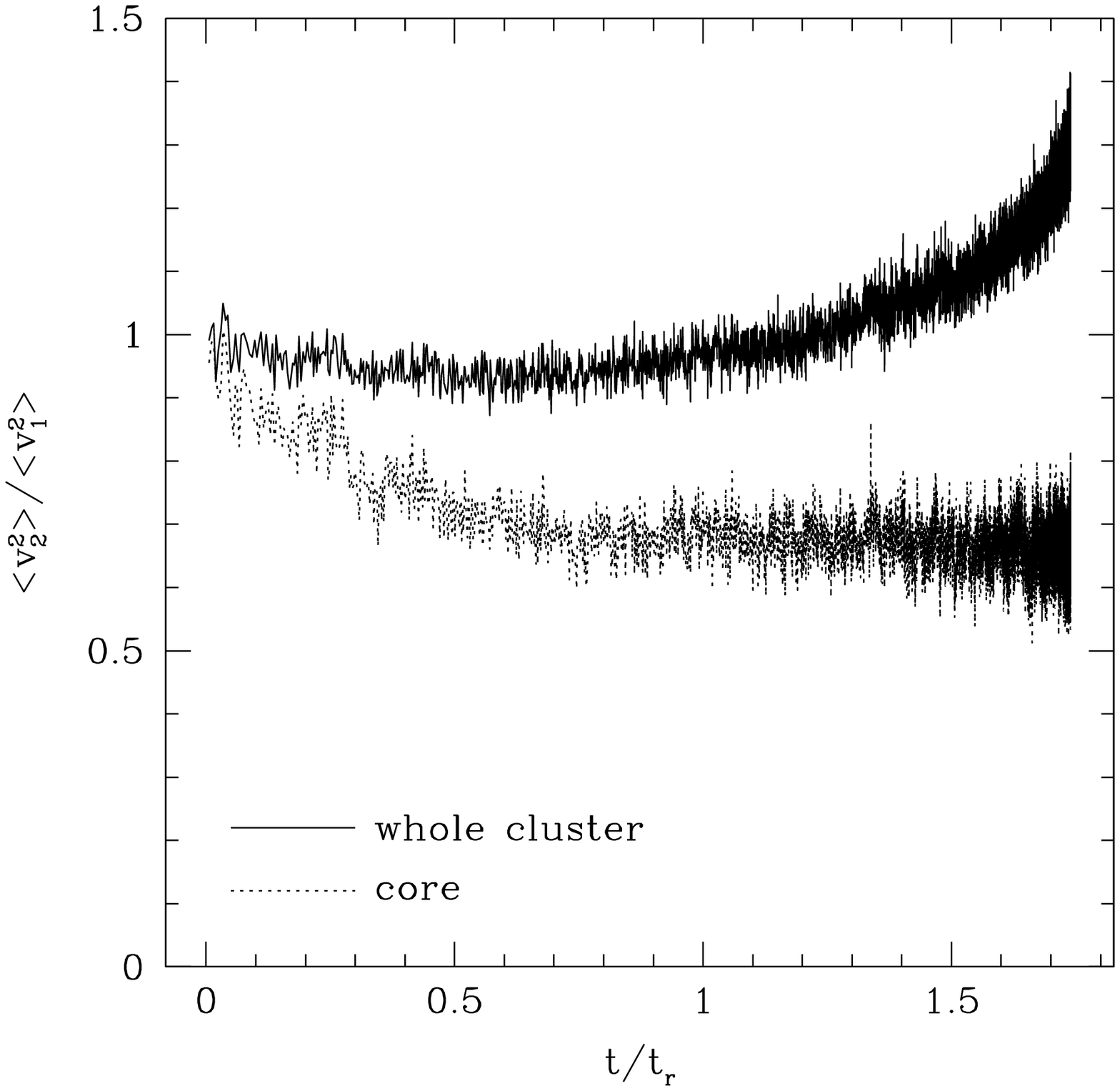}
\caption{\label{fig7}Ratio of rms velocities of the heavier component to the lighter component for the system shown
	in \fig{fig6}.  Solid line is this ratio for the cluster as a whole.  Note that this ratio remains 
	near unity for about the first half of the simulation.  For reference, the value of this ratio in the core 
	is shown as a dotted line.  This curve approaches $2/3$, consistent with equipartition.}
\end{inlinefigure}
\end{minipage}
\end{figure*}

\begin{figure*}[b]
\begin{minipage}[t]{0.47\linewidth}
\begin{inlinefigure}
\figurenum{8}
\plotone{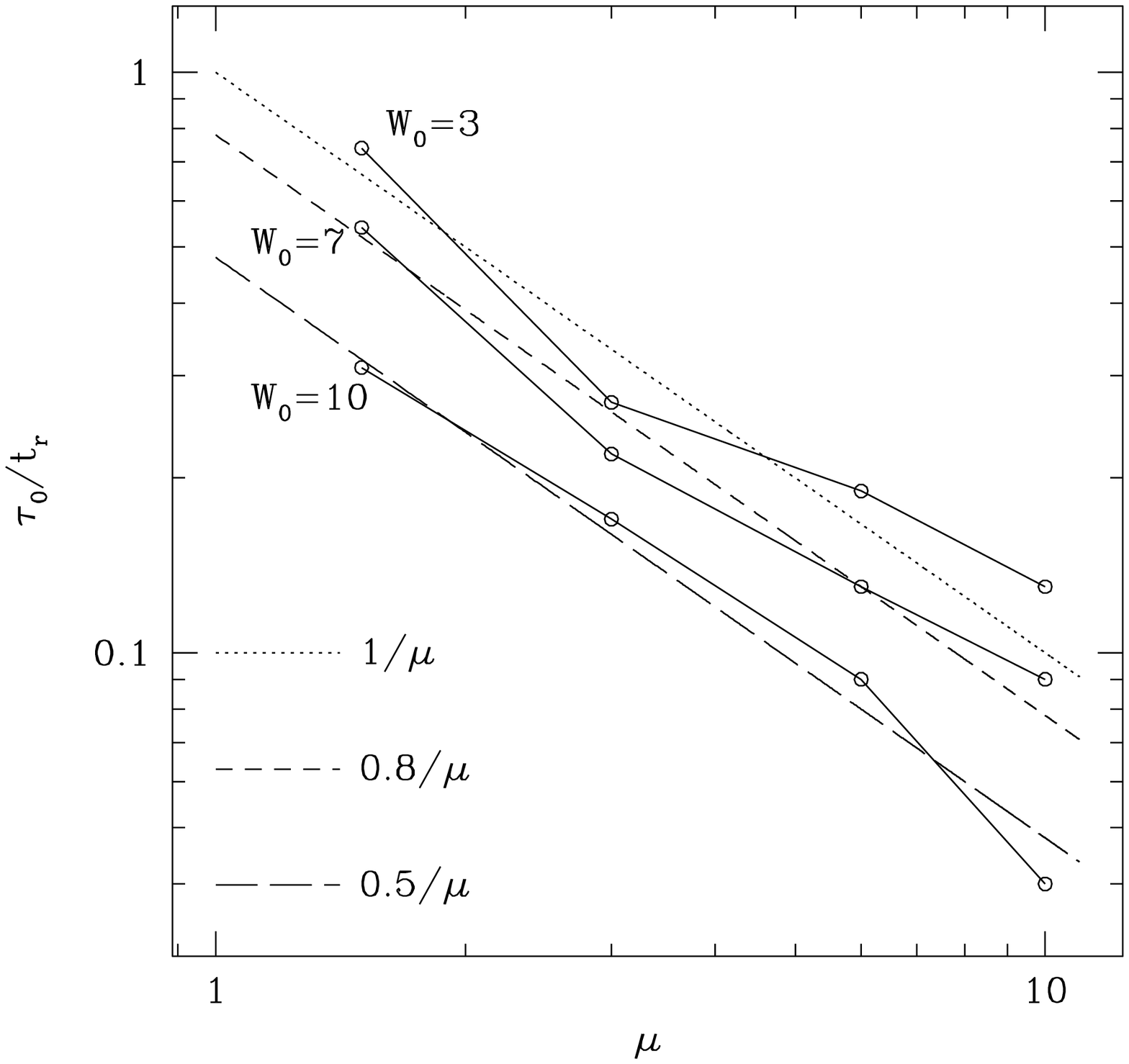}
\caption{\label{fig8}Mass-segregation timescale as a function of mass ratio, for a few King initial models.
	In all cases $\tau_0 \propto 1/\mu$, as predicted by simple theoretical
	arguments.}
\end{inlinefigure}
\end{minipage}
\hfill
\begin{minipage}[t]{0.47\linewidth}
\begin{inlinefigure}
\figurenum{9}
\plotone{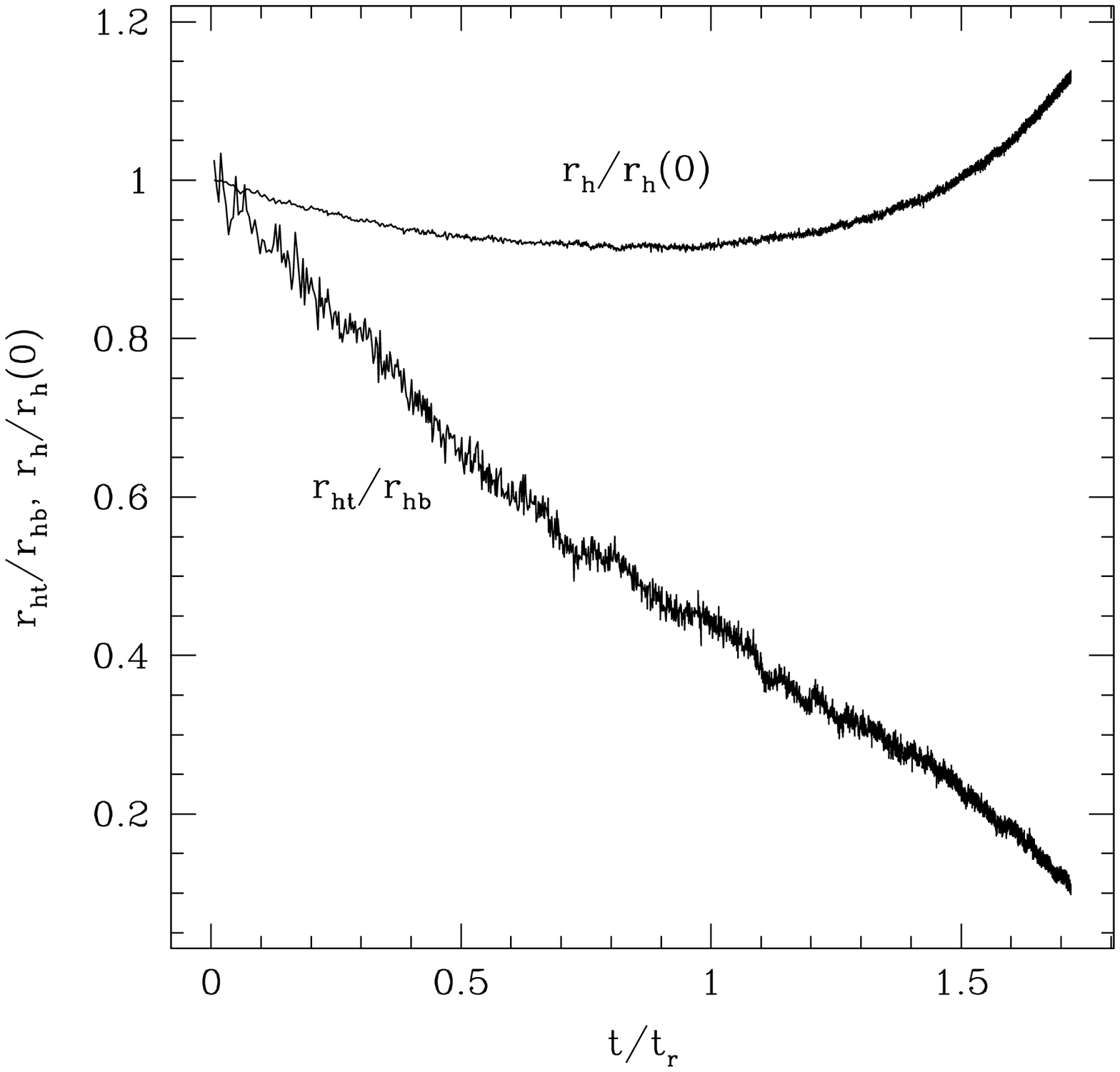}
\caption{\label{fig9}Ratio of half-mass radii of the tracers to the background stars for the system shown in 
	\fig{fig6}, displaying persistent segregation.  Also plotted, for reference, is the overall half-mass radius 
	of the cluster.}
\end{inlinefigure}
\end{minipage}
\end{figure*}

\section{Heavy Tracers}

\subsection{Analytic Results}

\citet{spitzer69} pioneered the theoretical analysis of mass segregation in clusters.  
Considering a two-component cluster in which one component is more massive than the other,
yet sparse enough to be considered a tracer population, he found that the timescale on 
which the heavier species contracts varies approximately as $1/\mu$.  For completeness, we include
a brief summary of the steps leading to this result.

For the generic case of energy transfer when $E_2$, the mean kinetic energy of the massive stars, is
greater than the corresponding energy $E_1$ for the lighter stars, the rate of energy loss of the 
massive stars is given by
\begin{equation}
\label{seg1}
{dU_2 \over dt} = {\left(E_1 - E_2\right) \over t_{\rm eq}} \, ,
\end{equation}
where $U_2$ is the mean total energy of the heavier species.  
The equipartition time is approximately \citep{spitzer40,spitzer62}
\begin{equation}
\label{seg2}
t_{\rm eq} = {\left(\langle v_1^2 \rangle+\langle v_2^2 \rangle\right)^{3/2} \over 
	8\left(6\pi\right)^{1/2} \rho_{01} G^2 m_2 \ln{N_1}} \, ,
\end{equation}
where $\rho_{01}$ is the central density of the lighter species, $N_1$ is the number of light stars
in the cluster, and brackets denote an average over stars.  When one expresses this timescale
in units of $t_{\rm r1}$, the overall relaxation timescale of species 1, one obtains
\begin{equation}
\label{seg3}
{t_{\rm eq} \over t_{\rm r1}} = {3 \pi^{1/2} \over 16} {m_1 \over m_2} 
	\left( 1 + {\langle v_2^2 \rangle \over \langle v_1^2 \rangle}\right)^{3/2} \, .
\end{equation}
It follows that if the rms velocities of the two species are equal, as they are at the start 
of our simulations (they remain roughly equal for the first half of the simulation; 
see \fig{fig7}), $t_{\rm eq}/t_{\rm r1}$ initially varies approximately as $1/\mu$.

To see how this timescales manifests itself, assume, for the sake of simplicity,
 that the kinetic energy of the heavier species,
$E_2$, dominates that of the lighter species, $E_1$.  \Eq{seg1} then becomes
\begin{equation}
\label{seg4}
{dU_2 \over dt} = {-E_2 \over t_{\rm eq}} \, .
\end{equation}
Application of the usual virial theorem yields $U_2 = -E_2 = {1 \over 2} W_2$, where $W_2$ is the potential
energy of species 2, and so we have
\begin{equation}
\label{seg5}
W_2 \propto e^{t/t_{\rm eq}} \, .
\end{equation}
Under the assumption that the potential of species 2 dominates, $W_2$ is proportional to 
$GM_2^2/r_{\rm 2h}$, where $r_{\rm 2h}$ represents any characteristic radius of species 2, 
but for concreteness has been chosen to be the half-mass radius.  Thus,
\begin{equation}
\label{seg6}
r_{\rm 2h} \propto e^{-t/t_{\rm eq}} \, ,
\end{equation}
which simply states that the timescale for contraction of the heavier species is 
given by $t_{\rm eq}$, the equipartition time.  \Eq{seg6} still holds true 
even when the potential of species 2 is not the dominant potential.

\subsection{Numerical Results}
For studying mass segregation of heavy tracers, we use 
as the unit of time the overall cluster relaxation time, $t_r$, given by 
\eq{time2}.  This time differs from $t_{\rm r1}$ (given by equation \ref{time3}), 
the relaxation time of species 1 and the unit of time in
\eq{seg3}, by a factor close to unity that varies slowly among sets of
initial conditions.  The quantity $t_r$ also differs from the
half-mass relaxation time, $t_{\rm rh}$ (given by equation \ref{time4}), 
the unit of time used in the
preceding sections for studying light tracers.  A good rule of thumb
is that $t_{\rm rh} \approx 0.1 t_r$ \citep{joshi1}.

To determine the mass segregation timescale---which should
approximately equal $t_{\rm eq}$ from \eq{seg6}, but which we label
$\tau_0$---we look at the number of tracers in the halo, defined to be
the region between the half-mass radius and the tidal radius, as a
function of time and fit to it a decaying exponential.  \fig{fig6}
shows a representative determination of this timescale for a $W_0=3$,
$\mu=1.5$, $N_t=1000$ model.  Note that the exponential fit is only
meaningful for the first few half-mass relaxation times, during which
the ratio of velocity dispersions $\langle v_2^2 \rangle / \langle
v_1^2 \rangle$ doesn't change appreciably.  \fig{fig7} shows this
ratio hovering at unity for nearly the first half of the simulation.
(For reference, $\langle v_2^2 \rangle / \langle v_1^2 \rangle$
averaged over the core is shown as a dotted line.  This curve
approaches 2/3, consistent with thermal equilibrium.)  \fig{fig8}
shows $\tau_0$ as a function of $\mu$ for King initial models with
$W_0=3$, 7, and 10.  In all cases, the mass segregation timescale
shows a clear $1/\mu$ dependence, as expected from \eq{seg3}.
\fig{fig9} gives another look at the segregation, showing the ratio of
half-mass radii of the tracers to the background stars.

Table 2 gives relevant initial conditions for all the heavy tracer
systems considered here, as well as the thermal equilibrium stability
parameters $S$ and $\Lambda$ (see eqs.~(\ref{two}) and~(\ref{three})), 
core collapse times, and the ratio of half-mass radii
for the two species at core collapse.  Similarly to Table 1 we have
included the following astrophysically relevant timescales: $\tau_{0}
(r_h<r<r_t)$, the $e$-folding timescale shown in \fig{fig8};
$\tau_{1.5} (r_c)$, the time for the number of heavy stars within the
initial core radius to increase by a factor of 1.5; $\tau_{10} (r_c)$,
the time for the number of heavy stars within the initial core radius
to increase by a factor of 10; and $\tau_{1.5} (r_h)$, the time for
the number of heavy stars within the initial half-mass radius to
increase by 1.5.  Here again, the trend is clear: the closer the mass
ratio is to unity, the longer it takes the heavy tracers to sink into
the core.  

The shorter core collapse time scales for the models
with larger $\mu$ or larger $W_0$ give these clusters little 
time to evolve dynamically.  Consequently, mass segregation is not 
as pronounced in these clusters, as is evident from Table 2.

\section{Summary and Discussion}

We have studied the mass segregation of both light and heavy tracers
in two-component star clusters.  The calculations were performed using our
recently developed Monte Carlo code using $N=10^5$ stars, 
with a few key calculations repeated using a direct {\nbody\/} code 
running on GRAPE computers.  The two methods showed good agreement in 
all cases.  The large $N$ that can be used in Monte Carlo 
simulations is essential for this type of study, as it allows us to treat
a small tracer population of objects without suffering 
large numerical noise.

We found that light tracers with $\mu \lesssim 0.1-0.4$ (depending on cluster
parameters) are ejected from the cluster
core, on average, within one central relaxation time ($\sim10^7-10^9\,$yr for most 
globular clusters).  The cores of globular clusters should therefore be
largely devoid of stars of mass $\lesssim 0.25\,M_\odot$. As a result, we
predict that any observed low-mass star in a dense globular cluster core
is most probably part of a binary system with a more massive companion.
For example, low-mass helium white dwarfs observed in a cluster core are
most probably in binary systems (Taylor et al.\ 2001).

Some of the low-mass objects, when they are ejected from the central
region of the cluster, settle in the outer halo, where the relaxation time
is so long that they are prevented from further segregation or evaporation
through the tidal boundary. We found in
our simulations that the
number density of light tracers in the outer parts of the cluster can
actually become significantly higher than it was initially. We studied
the implications of these results for gravitational microlensing in
globular clusters and found that low-mass objects in the cluster halo
can dominate the optical depth for microlensing by
up to an order of magnitude. We applied these results
to the recent null detection of \citet{sahu2} to
estimate an upper limit on the current mass
fraction of M22 in very low-mass objects
of $\sim 25\%$ at the 63\% confidence level.

For clusters with heavy tracers ($\mu > 1$) we found good
agreement between our Monte Carlo results and direct 
{\nbody} integrations, as well as simple theoretical
estimates.  In particular, we found that the time scale for mass segregation, $\tau_0$,
varies as $1/\mu$, as expected from theoretical predictions.
Specifically, over a wide range of initial cluster concentrations and mass ratios,
$\tau_0/t_r = C/\mu$, where the constant $C\simeq 0.5-1$.

\acknowledgements 
We are grateful to Eric Pfahl and Saul Rappaport for helpful
discussions and comments.  SPZ is grateful to Northwestern 
University for their hospitality.  This work was supported by NSF Grant
AST-9618116 and NASA ATP Grant NAG5-8460. SPZ was supported by 
Hubble Fellowship grant HF-01112.01-98A, awarded by the Space Telescope Science
Institute, which is operated by the Association of Universities for
Research in Astronomy, Inc., for NASA under contract NAS\,
5-26555. Our Monte Carlo simulations were performed on the Cray/SGI
Origin2000 supercomputer at Boston University under NCSA Grant
AST970022N.  The direct {\nbody\/} simulations were performed on the GRAPE
systems at the University of Tokyo and at Drexel University.

\end{document}